\documentclass[12pt]{article}
\usepackage{arxiv}
\usepackage{latexsym,amsmath,amssymb,amsbsy,graphicx,bm} 
\usepackage[T1]{fontenc}
\usepackage[cp1251]{inputenc} 
\usepackage[font=small,labelfont=bf]{caption}

\def\bsr{\boldsymbol{\rho}}
\def\bsa{\boldsymbol{\alpha}}

\title{The self-learning AI controller for adaptive power beaming with fiber-array laser transmitter system}

\author{
				{A.M. Vorontsov}, \\
				Kaspersky Lab USA,\\
				500, Unicorn Park, Woburn, MA, 01801, USA,\\
				\textit{Artem.Vorontsov@kaspersky.com}\\		
				\And
				{G.A. Filimonov}, \\
				Intelligent Optics Laboratory, School of Engineering, University of Dayton, \\
				Dayton, Ohio, 45469, USA,\\
				\textit{gfilimonov1@udayton.edu}
		    }
				
\renewcommand{\shorttitle}{\textit{arXiv} Template}

\begin{document}

\maketitle

\begin{abstract}
 
In this study we consider adaptive power beaming with fiber-array laser transmitter system in presence of atmospheric turbulence. For optimization of power transition through the atmosphere fiber-array is traditionally controlled by stochastic parallel gradient descent (SPGD) algorithm where control feedback is provided via radio frequency link by an optical-to-electrical power conversion sensor, attached to a cooperative target. The SPGD algorithm continuously and randomly perturbs voltages applied to fiber-array phase shifters and fiber tip positioners in order to maximize sensor signal, i.e. uses, so-called, ``blind'' optimization principle.

In opposite to this approach a perspective artificially intelligent (AI) control systems for synthesis of optimal control can utilize various pupil- or target-plane data available for the analysis including wavefront sensor data, photo-voltaic array (PVA) data, other optical or atmospheric parameters, and potentially can eliminate well-known drawbacks of SPGD-based controllers. In this study an optimal control is synthesized by a deep neural network (DNN) using target-plane PVA sensor data as its input. A DNN training is occurred online in sync with control system operation and is performed by applying of small perturbations to DNN's outputs. This approach does not require initial DNN's pre-training as well as guarantees optimization of system performance in time. All theoretical results are verified by numerical experiments. 
                  
\end{abstract}


\section{Introduction.}
\label{s0}

The interest in development of fiber-array laser transmitter systems \cite{LeNiHu, KuKaLe} has been steadily growing over the past two decades due to its compact size and low cost of system components comparing with conventional systems having the same transmitting aperture size. The presence of adaptive beam control system makes fiber-arrays capable for direct integration and digital implementation of exotic beam shaping \cite{ExBeam}, adaptive mitigation of the propagation-medium-induced phase aberrations, target tracking and beam pointing. All this allows to consider fiber-array as potential laser transmitter system for a wide class of optical applications -- from directed energy to free-space optical communications \cite{KuKaLe, VoFiOv}, additive manufacturing \cite{FiKoVo} and power beaming \cite{TriGo}.

A wireless delivering energy to remote and hard-to-reach power consumers with laser transmitters is promising technology especially taking into account explosive growth number of rechargeable portable devices and equipments. In recent power beaming experiments with single-aperture laser transmitters charging of remotely located mobile devices including drones \cite{AsStuGur}, vehicles \cite{JinZhou} and even cell phones \cite{KaMaTu} was performed where for optical-to-electrical power conversion photo-voltaic array (PVA) panels attached to a particular device were used. As far as all mentioned experiments occurred in lower Earth atmosphere authors especially reported significant dependence of laser energy transfer efficiency on atmospheric conditions along a beam propagation path. By this reason in \cite{JinZhou} only 25\%\footnote{This estimation includes PVA transfer energy efficiency also.} of transmitted laser energy was delivered to a mini rover and in \cite{KaMaTu} for increasing of energy transfer efficiency authors optimized laser beam position trying to keep it near PVA center. Thus, in realistic conditions appropriate power beaming efficiency can be achieved using adaptively controlled laser systems only including fiber-arrays.  

\begin{figure}[hbtp]
\centering
\includegraphics[width=0.8\textwidth]{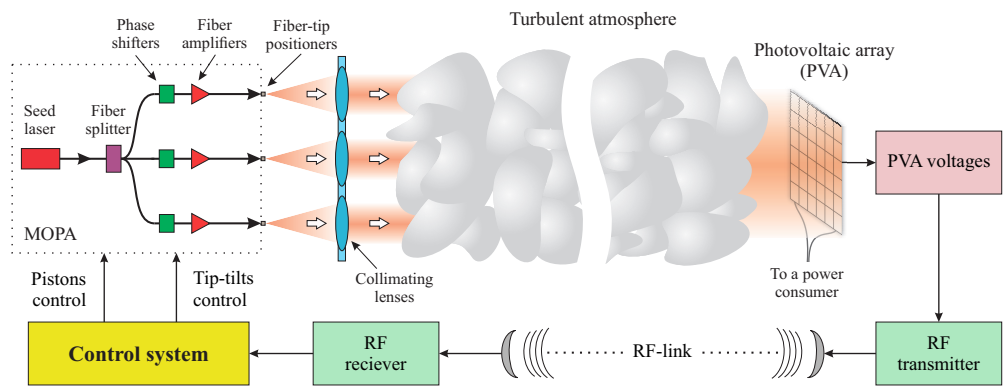}
\caption{Notional schematics of the fiber-array transmitter system in the problem of power beaming through turbulent atmosphere.} \label{fig_01}
\end{figure}

Consider the problem of optimal energy transfer through the atmosphere onto remote high-resolution PVA panel. Let for mitigation of the atmospheric-induced received power losses from a non-optimal match between photovoltaic conversion and the projected laser beam footprint, errors in laser beam pointing, turbulence-induced beam spread, wander and distortion controlled fiber-array laser transmitter system is used. Suppose that conformal laser beam outgoing from an array of optical collimators passes through the turbulent atmosphere and illuminates PVA (see Fig. \ref{fig_01}). The PVA performs optical-to-electrical power conversion with delivering the transferred energy to a power consumer and in addition has a radio-frequency (RF) link with multi-channel master oscillator/power amplifier (MOPA) fiber system. This link allows rapidly (in comparison with turbulence changing time) to transfer PVA grid power values directly to the control system where this information can be used for optimization of phase shifts and fiber tip positions. The control efficiency in power beaming is traditionally measured using several well-known indicators (or metrics) including overall transferred energy magnitude and a parameter responsible for matching optimal power distribution on PVA.

The most common fiber-array optimization controller uses stochastic parallel gradient descent (SPDG) algorithm for syntheses of the control. Optimization of metric value by SPGD algorithm is continuously performed by generating and applying of small random perturbations to fiber-array actuators. If current perturbation increases (for metric maximization problem) metric value, the optimization step is considered as completed, otherwise, depending on algorithm type, a new perturbation is generated or existing perturbation is applied with an opposite sign. Modern SPGD controllers used for fiber arrays coherent combining allow to perform up to $5\times10^{5}$ parallel optimization steps per second that provides desirable performance for a wide range of fiber-array configurations, propagation scenarios and atmospheric conditions. However this technique has several well-known drawbacks: (1) SPGD convergence is rapidly decreasing with increase of the number of subapertures above 50-100, (2) SPGD optimization algorithm is extremely sensitive to perturbation statistics and type of gain coefficient that in each particular case are chosen empirically, (3) SPGD realizes, so-called, ``blind'' optimization strategy and ignores any additional information available for the analysis and potentially useful for synthesis of the control, for example, PVA power distribution as in the considered case.

One of the most promising way to avoid aforementioned drawbacks is an enhancing of SPGD optimization algorithm using state-of-the-art deep learning (DL) methodology. The DL paradigm supposes to utilize deep neural networks (DNN) of various types and topologies for extraction and analyzing of information about interaction of the control system with environment in order to synthesize optimal control. Using of DNN in a control system supposes to have some training mechanism and here two baseline approaches can be proposed: 
(a) offline training when the controller's DNN is specially pretrained in offline regime and only than is applied for real-time control purposes and (b) self-learning when randomly initiated DNN is continuously interacting with changing environment is trained on-the-fly using performance metrics as stimulus for learning and observed environment characteristics as the input \cite{Wat}. 

There are numerous papers (see, for example, \cite{WaDuZh, GZM} and literature in these papers) dedicated to coherent beam combining using DNN with offline training strategy. Methods of DNN utilization for syntheses of optimal control are significantly varied for different authors and include direct generation of control by DNN using target-plane measurements \cite{LoGl}, generation by DNN of some initial pupil-plane phase distributions \cite{MGP} and more sophisticated cascaded schemes where DNN is operated in pair with SPGD \cite{GuLiLi}. However, this methodology has also several important drawbacks: (1) offline training technique supposes to manually collect or simulate huge training datasets that should cover all potential system application scenarios, (2) one should guarantee that optical system will always operate in the same environmental conditions as represented in this training set, so that any extension of condition ranges or any system modification automatically requires to supplement (in the best case) or completely rebuild (in the worst case) of training set with the following DNN retraining. In other words, the offline learning scheme does not support any system adaptation to any changes and primary suitable for the systems which are interacted with the statistically constant environments. 

In opposite, reinforcement learning (RL) approach \cite{Rlearn} initially is designed for operation in unknown and changing environment. The major idea of this methodology is to place completely untrained intellectual agent (controller, in our terminology) in some environment, to provide their continuous mutual interaction and to define some reward function that should be maximized during these interactions. Here, the reward function plays a role of stimulus to agent's learning and its adaptation to environment changes, stimulus to forget unusual or outdated information about the environment or about interaction with the environment, stimulus to follow optimal trajectories. The agent-environment interaction in RL is considered in terms of ``action-state'' space where agent's action is typically represented in the form of multidimensional control vector and for describing of environment state available and currently observed information (f.e., instantaneous sensor measurements, snapshot images, etc.) is used.   

The problem of fiber-array coherent beam combining can be classified in RL paradigm as ``multidimensional continuous action space'' problem that commonly considered using deep Q-learning \cite{qlearn2} approach or more precisely deep deterministic policy gradient (DDPG) algorithm \cite{DDPG}. In this approach two separate DNN are considered -- one a-DNN represents intellectual agent as well as other one q-DNN is responsible for simulation of environment response. Traditionally this response is chosen in the form of Bellman's $Q$-function \cite{Bell} so that q-DNN will approximate it during agent-environment interaction. The training process is synchronized for both DNNs -- an agent performs some (initially random) actions, explores environment's reaction and takes reward thereby training q-DNN that provides local approximation of environment response and allows to predict this response for future local actions. Built approximation of $Q$ function allows to take (symbolic) gradient over q-DNN's inputs, then to use this gradient for correction of a-DNN trainable weights in accordance with policy gradient theorem and hence to improve agent future actions.

The proposed DDPG algorithm can be obviously applied to considered power beaming problem \cite{KeXuXu} if, for example, as the reward function one takes aforementioned system performance metric, as intellectual agent consider beam controller and as environment state -- target-plane PVA intensity distribution and instantaneous metric magnitude. However, this straightforward approach has one important drawback -- in this case one factually needs to build DNN approximation of metric response to all current and following controller's actions (i.e. $Q$-function for this case). For optical systems operated in realistic conditions performance metric depends on numerous factors, metric reaction on piston/tip-tilt control is non-linear and can be changed in time, number of control channels is big and these channels are strongly coupled. In these conditions there is no chance to build robust, accurate and static approximation of $Q$-function that will automatically lead to poor training of controller's DNN and completely avoid all advantages given by RL.      
   
In this paper we introduce SPGD-based training procedure of DNN controller, where instead of consideration and differentiation of Bellman's $Q$-function we use SPGD-type estimation of environment's response gradient. This technique is especially suitable for rapidly changing non-linear environments and multidimensional control space specific for the problem of energy transferring through the atmosphere. The idea is to continuously pass all available incoming information about environment state (PVA intensity distribution, metric) and controller's action (actuator voltages) to DNN input in order to synthesize optimal control in these particular conditions and time moment. At the same time DNN outputs are continuously perturbed using SPGD-type procedure that provides self-learning of DNN as well as guarantee minimization of performance metric in time. In this paper this type of control will be called ``active AI control'' meaning that here simultaneously two independent processes are taking place -- ``active'' SPGD-type metric optimization with parallel DNN training and ``passive'' DNN inference with synthesis of the control.   

In Section \ref{s1} the problem statement, theoretical background of SPGD-based optimal control and active AI control will be given. In Section \ref{s2} implementation details for the proposed in Section \ref{s1} approach will be stated. In Section \ref{s3} training capabilities and performance of metric optimization for the proposed AI control system is demonstrated via numerical experiments.

\section{Basic considerations.}
\label{s1}
In this section we introduce the power beaming problem as well as basic principles of SPGD and active AI control. 

\subsection{The power beaming problem.}
\label{s1_1} 

\begin{figure}[hbtp]
\centering
\includegraphics[width=0.7\textwidth]{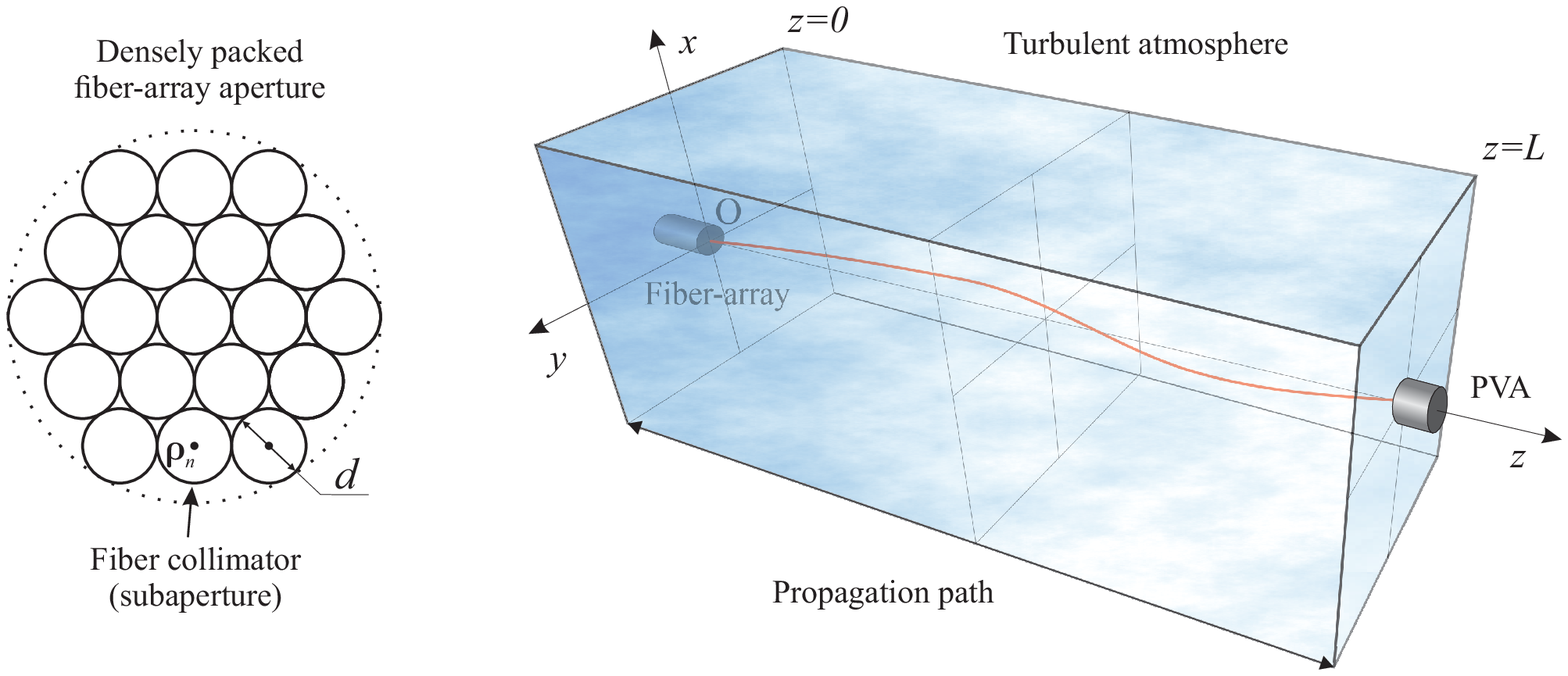}
\caption{Fiber-array aperture with $N_{\rm{sa}}=19$ subapertures (left) and geometry of optical field propagation (right).} \label{fig_02}
\end{figure}

Consider optical wave propagation along (parallel to) the axis $Oz$ of Cartesian system $Oxyz$, assuming that the fiber-array transmitter is located at the coordinate origin $O$ and PVA panel -- at the point $(0,0,L)$, where $L > 0$ is propagation distance, see Fig. \ref{fig_02} on the right. Let $\boldsymbol{r}=(x,y,z)$ be a vector in $Oxyz$, $\bsr=(x,y)$ is a plane coordinates and $t\ge 0$ denotes time. Propagation of linear polarized monochromatic optical waves with complex amplitude $U(\boldsymbol{r},t)=U(\bsr,z,t)$ through an optically inhomogeneous medium is commonly described by the following parabolic equation \cite{RyKraTat}:

\begin{equation}\label{Parab}
2ik\frac{\partial U(\bsr,z,t)}{\partial z}+
\nabla_{\bot}^{2}U(\bsr,z,t)+2{k}^{2}n(\bsr,z,t)U(\bsr,z,t)=0,
\end{equation}

where $k$ is the wave number, $\nabla_{\bot }^{2}=\partial^{2}/{\partial x}^{2}+\partial ^{2}/{\partial y}^{2}$ and $n(\bsr,z,t)=$ $n(\boldsymbol{r},t)$ is a random function corresponding to the refractive index fluctuations having zero meanvalue 
$\left< n(\boldsymbol{r},t) \right>=0$. The notation $\left<\cdot \right>$ is used to describe statistical averaging over the ensemble of refractive index realizations. The boundary conditions for complex amplitude $U$ at the plane $z=0$ are defined as:
\begin{equation*}
U(\bsr,z=0,t)=U_{0}(\bsr, t),
\end{equation*}

where $U_{0}(\bsr, t)$ is the optical field complex amplitude at the transmitter plane.

Assume that the fiber-array transmitter system has $N_{\rm{sa}}$ hexagonal subapertures (see Fig. \ref{fig_02}, left) and each particular subaperture is described by a stepwise function $H_n(\bsr) = H(\bsr - \bsr_n)$, $n\in \{ 1,..., N_{\rm{sa}}\}$, where $\bsr_n=(x_n, y_n)$ is coordinate of subaperture center and

\begin{equation*}
H(\bsr)=
\begin{cases}
1,\,\, |\bsr| \le d/2, \\
0, \,\,\text{otherwice},
\end{cases}
\end{equation*}

where $d>0$ is subaperture diameter. For the complex amplitude of optical field at the fiber-array output (pupil) plane $\{z=0\}$ one has:

\begin{equation}\label{eq01}
U_0\left(\bsr, t\right) = 
A_0 \sum\limits_{n = 1}^{N_{\rm{sa}}} {H_n\left( {\bsr} \right)}\,
e^{-{\left|{\bsr - \bsr_n} \right|^2}/{a_0^2}}\,
e^{ik\varphi_n \left(\bsr, t \right) },
\end{equation}

where $A_0\ge 0$ is the constant dependent on the power transmitted through the fiber-array and ${\varphi _n}\left(\bsr, t\right) = {\bf{s}}_{n}(t) \cdot \left(\bsr - \bsr_n \right) + c_n(t)$, $n \in \{1,...,N_{\rm{sa}}\}$ is the linear over $\bsr$ function representing phase component of the outgoing field in the $n$-th subaperture at time $t\ge 0$. Here $ {\bf{s}}_n(t) = \left(s_{n,x}(t), s_{n,y}(t)\right) \in \mathbb{R}^2$ and $c_n(t)\in\mathbb{R}$ are, respectively, tip-tilt and piston components of the $n$-th beamlet and dot denotes scalar product of two vectors. The beamlet radius $a_0$ corresponds to $e^{-1}$ fall-off in intensity. 

Adaptive beam shaping and compensation of turbulence-induced aberrations can be performed using control of either solely piston $\left\{c_n(t)\right\}_{n = 1}^{N_{\rm{sa}}}$ or piston and tip-tilt $\left\{ {\bf{s}}_n(t) \right\}_{n = 1}^{N_{\rm{sa}}}$ phase components. For future considerations denote as $K$ the number of all control parameters so that $K=N_{\rm{sa}}$ for single piston control (then $s_{n,x}(t)$ and $s_{n,y}(t)$ are constants for any $n \in \left\{1,...,N_{\rm{sa}} \right\}$ and $t \ge 0$) or $K=3N_{\rm{sa}}$ for both piston and tip-tilt control. 

\subsection{Performance metric for power beaming problem.}
\label{s1_2} 

Now suppose that considered fiber-array transmitter is equipped by a control system operating with aforementioned phase components. The control performance is traditionally estimated using several well-known metrics suitable for particular transfer energy task. For example, overall fiber-array transmitted power is accounted using integral power-in-the-bucket (PIB) metric, for a beam shaping problem this metric should be accompanied with the term, accounting power distribution on PVA, for directed energy applications Strehl ratio dedicated to power peak-to-valley measurement is commonly used. The PIB and Strehl ratio can be measured directly using photo-voltaic sensors or computed using target-plane intensity distribution $I\left({\bsr,z = L,t} \right)=\left| U \left({\bsr,z = L,t} \right)\right|^2$ in case if PVA is attached to the target.

In this study we will suppose that at the target-plane $z=L$ there is a high-resolution square PVA centered at the point $(0,0,L)$ with sides having the size $D>0$ and parallel to $x$ and $y$ axes. As the performance metric we will consider, so-called, smooth Strehl ratio: 

\begin{equation}\label{J}
J_t = J(t) =
\frac{1}{J_{\rm{vac}}}
\int{I \left(\bsr,z=L,t \right)
 e^{-\left| \bsr \right|^2/\beta^2}
d^2 \bsr},
\end{equation}

where $\beta >0$ is smoothing parameter and $J_{\rm{vac}}$ is normalization factor for providing the condition $J_t\le1$ for any $t\ge0$. This condition will be guaranteed if we consider free-space propagation of the initial complex field $U_{0}$ with $n(\bsr,z,t)\equiv 0$ in \eqref{Parab}, take corresponding target-plane intensity distribution $I_{\rm{vac}} \left(\bsr, z=L, t \right)$ and compute $J_{\rm{vac}}$ as:

\begin{equation}\label{Jvac}
J_{\rm{vac}} = \max \left\{
\int{I_{\rm{vac}}\left(\bsr,z=L,t \right)
 e^{-\left| \bsr \right|^2/\beta^2}
d^2 \bsr}\right\},
\end{equation}

where maximum is taken over all $K$ control parameters. Note that all target-plane intensity distributions are measured on PVA (see Fig. \ref{fig_01}) so that integration in \eqref{J} and \eqref{Jvac} is also limited to PVA area only.

\subsection{Basic principles of SPGD control.}
\label{s1_3} 

Denote the entire set of control variables as ${\bf{u}} = \left( {u^1,...,u^K} \right)$. Consider general case when performance metric depends on a vector of some pupil- or target-plane field characteristics 
${\bf{I}}_t = {\bf{I}}_t\left(\bsr;{\bf{u}} \right) = \left(I^1\left( \bsr, t;{\bf{u}} \right),...,I^M\left(\bsr, t;{\bf{u}} \right) \right)$, $M\ge1$, where each particular characteristic, in turn, depends on control parameters as well as can be considered as spatial and time dependent function. This notation covers metric dependence on one- or two-dimensional field characteristics such as time series, intensity and phase distributions. For example, for considered above smooth Strehl ratio one has $M=1$ and $I^1\left( \bsr, t;{\bf{u}} \right) = I\left({\bsr,z = L,t} \right)$ is time-dependent target-plane intensity image and for conventional Strehl ratio $I^1\left(t;{\bf{u}} \right) = I\left({\bsr={\bf{0}},z = L,t} \right)$ is time series. 

Then, for the performance metric one has: 
\begin{equation*}
J_t = 
J_t\left( {\bf{u}} \right) = J\left( {{{\bf{I}}_t}\left( \bsr;{{\bf{u}}} \right)} \right)
\end{equation*}

and the problem of synthesis of optimal control can be formulated as follows:

\begin{equation}\label{eq02}
{J_t}\left( {\bf{u}} \right) \to \mathop {\max }\limits_{\bf{u}}.
\end{equation}

The optimization problem \eqref{eq02} can be numerically solved using well-known gradient descent optimization algorithm. Let an optimal control trajectory be 
${\bf{u}}_t = {\bf{u}}\left( t \right) = \left({u_t^1,...,u_t^K}\right) = \left( {u^1\left(t \right),...,u^K\left(t \right)} \right)$. 
Taking as the time increment the unity step the iteration process will have the form: 

\begin{equation}\label{eq03}
u_{t + 1}^k = u_t^k + {\gamma _t}\frac{{\partial {J_t}}}{{\partial {u^k}}}d{u^k}, \,\,\, k \in \left\{ {1,...,K} \right\},
\end{equation}

where $\gamma _t^{} = \gamma \left( {{J_t},t} \right)$ is time-dependent gain coefficient and $\left\{ {d{u^k}} \right\}_{k = 1}^K$ are positive scaling factors corresponding to different control sensitivities. In case when metric gradient over control parameters can be derived analytically or numerically the algorithm \eqref{eq03} (or its numerous variations), as the rule, provides rapid and robust computation of optimal trajectory's approximation. However, for real-world physical systems when $J_t\left( {\bf{u}} \right)$  is measured in real-time regime and computation of metric gradient is performed via applying component-wise perturbations to control variables the iteration process \eqref{eq03} has important drawback. The point is that with increasing of the number of control parameters and under assumption of sufficiently fast changing of physical conditions (f.e., atmospheric turbulence is changing in approximately 1 ms) the calculation of gradient in \eqref{eq03} will be lag behind environmental changes and as the result the gradient approximation will be inaccurate. 

The SPGD algorithm \cite{Spall, SPGD} is dedicated to avoid this problem. The major idea of SPGD optimization is in fast estimation of metric gradient using simultaneous perturbations of all control parameters. Let ${\bf{u}}_{t + 1} = {\bf{u}}_t + \delta {\bf{u}}_t$, where $\delta {\bf{u}}_t = \left( {\delta u^1,...,\delta u^K} \right)$ is a small increment that can be applied to controls in each time step. Then $J_{t + 1} = J\left( {{\bf{I}}_{t + 1}\left(\bsr;{{\bf{u}}_{t + 1}} \right)} \right) = {J_t} + \delta {J_t}$ and

\begin{equation}\label{eq04}
\delta {J_t} = 
{J_{t + 1}} - {J_t} \approx \sum\limits_{k = 1}^K {\frac{{\partial {J_t}}}{{\partial {u^k}}}\delta u_t^k}  + 
\sum\limits_{m = 1}^M {\frac{{\partial {J_t}}}{{\partial {I^m}}}\frac{{\partial I_t^m}}{{\partial t}}}. 
\end{equation}

Under assumption of relatively small changing of environment characteristics ${\bf{I}}_t$ per one time step in comparison with their reaction on the control perturbation, i.e. if:

\begin{equation*}
\left| \frac{\partial I_t^m}{\partial t} \right| \ll 
\left| {\sum\limits_{k = 1}^K {\delta u_t^k\frac{\partial I_t^m}{\partial u^k}} } \right|, \,\,\, m \in \left\{ {1,...,M} \right\},
\end{equation*}

the second term in \eqref{eq04} can be omitted so that one has:

\begin{equation*}
\delta {J_t} \approx \sum\limits_{k = 1}^K {\frac{{\partial {J_t}}}{{\partial {u^k}}}\delta u_t^k}.
\end{equation*}

Suppose that applied perturbations have stochastic nature and $\delta {\bf{u}}_t$ is a random vector with the following statistical properties:    

\begin{equation}\label{eq_05}
\begin{gathered}
\left\langle {\delta u_t^k} \right\rangle  = 0, \,\,\, k \in \left\{ {1,...,K} \right\}, t\ge0,\\
\left\langle {\delta u_t^k \delta u_t^l} \right\rangle  = B_t^{kl}, \,\,\, k,l \in \left\{ {1,...,K} \right\}, t\ge0,\\
\left\langle {\delta u_t^k \delta u_{t'}^l} \right\rangle  = 0, \,\,\, \text{for any} \,\,\, k,l \in \left\{ {1,...,K} \right\} \,\,\, \text{and} \,\,\, t\neq t', \,t, t'\ge0,
\end{gathered}
\end{equation}

where $\left\langle \cdot \right\rangle$, as before, denotes statistical averaging under ensemble of realizations, the diagonal correlation matrix $B_t^{} = \mathop {{\rm{diag}}}\limits_{k = 1,...,K} \left\{ {{{\left( {\sigma _t^k} \right)}^2}} \right\}$ has the size $K \times K$ and $\sigma _t^k > 0$, $k \in \left\{ {1,...,K} \right\}$ are standard deviations which will be specified below. Then, taking into account \eqref{eq_05}, one has for $l \in \left\{ {1,...,K} \right\}$: 

\begin{equation*}
\left\langle {\delta {J_t}\delta u_t^l} \right\rangle  \approx 
\left\langle {\sum\limits_{k = 1}^K {\frac{{\partial {J_t}}}{{\partial {u^k}}}\delta u_t^k\delta u_t^l} } \right\rangle  = 
\frac{{\partial {J_t}}}{{\partial {u^l}}}\left\langle {{{\left( {\delta u_t^l} \right)}^2}} \right\rangle  = 
\frac{{\partial {J_t}}}{{\partial {u^l}}}{\left( {\sigma _t^l} \right)^2}.
\end{equation*}

Changing in the last formula statistical averaging $\left\langle \cdot \right\rangle$ by averaging over time and taking into account aforementioned assumption about contribution of control and environmental factors one can rewrite the iteration process \eqref{eq03} in the form:

\begin{equation}\label{eq_06}
u_{t + 1}^k = u_t^k + \frac{{{\gamma _t}}}{{\sigma _t^k}}\delta {J_t}\delta u_t^k, \,\,\, k \in \left\{ {1,...,K} \right\}.
\end{equation}

The formula \eqref{eq_06} represents canonical SPGD optimization (maximization) algorithm which implementation will be discussed in details in Section \ref{s2_1}. 

The SPGD control algorithm provides appropriate optimization for a wide range of fiber-array configurations, performance metrics and atmospheric conditions. The computation of metric gradient estimation by SPGD algorithm is simple and can be performed extremely fast so that novel SPGD controllers can achieve the performance up to $5\times10^{5}$ iterations per second. All these facts motivate us try to adapt SPGD gradient estimation not solely for metric optimization but for DNN training purposes too.

\subsection{Basic principles of active AI control.}
\label{s1_4} 

From the consideration above it is clear that SPGD controller uses ``blind'' optimization principle when metric maximization is performed using solely metric perturbations and any additional and potentially useful information for solving the optimization problem is ignored. In opposite to this approach an AI controller have to be designed in such a way that it could collect and analyze all available information about physical process in order to synthesize optimal control potentially faster and more robust than SPGD. 

Recall that in our considerations this additional information is represented via the vector ${{\bf{I}}_t} = {{\bf{I}}_t}\left(\bsr;{{\bf{u}}} \right)$ of observed physical characteristics as well as their historical values. Hence, in order to account this information for the synthesis of optimal control the vector ${\bf{u}}_t$ now should be represented as a function dependent on instantaneous and retrospective values of ${\bf{I}}_{\tau}$ and ${J_{\tau}}$ taken for the time $\tau \le t$ and ${\bf{u}}_{\tau}$ taken for $\tau < t$. On the other hand, in order to make the controller capable for learning, i.e. capable for analyzing and correct utilizing the incoming information one has to include into this functional dependence a set of trainable parameters for their adjusting during controller operation. So, AI controller should have the form:

\begin{equation*}
{\bf{u}}_t = 
{\bf{U}}\left( {{{\bf{I}}_{\tau \le t}},{{\bf{u}}_{\tau < t}},{J_{\tau \le t}};\bsa} \right),
\end{equation*}

where ${{\bf{U}}_t}\left(\bsa\right) = {\bf{U}}\left( {{{\bf{I}}_{\tau \le t}},{{\bf{u}}_{\tau < t}},{J_{\tau \le t}};\bsa} \right) = \left\{ {{U^k}\left( {{{\bf{I}}_{\tau \le t}},{{\bf{u}}_{\tau < t}},{J_{\tau \le t}};\bsa} \right)} \right\}_{k = 1}^K$,  is some unknown vector function and $\bsa = \left( {{\alpha _1},...,{\alpha _P}} \right)$, $P \ge 1$ is the vector of optimization (trainable) parameters. For the future considerations denote $U_t^k = U_t^k\left( \bsa \right) = {U^k}\left( {{{\bf{I}}_{\tau \le t}},{{\bf{u}}_{\tau < t}},{J_{\tau \le t}};\bsa} \right)$. Then, the optimization problem \eqref{eq02} can be rewritten as:

\begin{equation}\label{eq_07}
\begin{gathered}
J_t = J_t \left(\bsa \right) = J\left( {\bf{I}}_t \left( \bsr; {\bf{U}}_t \left( \bsa \right) \right) \right),\\
J_t \left( \bsa \right) \to \max\limits_{\bsa}. 
\end{gathered}
\end{equation}

As before, the gradient descent optimization algorithm immediately gives us the following iteration process for the parameters $\bsa$:

\begin{equation*}
\alpha _{t + 1}^p = 
\alpha _t^p + 
{\gamma _t}\sum\limits_{k = 1}^K {\frac{{\partial {J_t}}}{{\partial {u^k}}}\frac{{\partial U_t^k}}{{\partial {\alpha ^p}}}d{\alpha ^p}}, \,\,\, p \in \left\{ {1,...,P} \right\}, 
\end{equation*}

where $\gamma_t = \gamma \left( {{J_t},t} \right)$ is the gain coefficient and $\left\{ {d{\alpha ^p}} \right\}_{p = 1}^P$ are positive scaling factors.  The estimation of gradients $\left\{ {\partial {J_t}/\partial {u^k}} \right\}_{k = 1}^K$ in the last formula can be performed as before using random perturbations. Then: 

\begin{equation}\label{eq_08}
\alpha _{t + 1}^p = 
\alpha _t^p + {\gamma _t}\delta {J_t}\sum\limits_{k = 1}^K {\frac{{\delta u_t^k}}{{\sigma _t^k}}\frac{{\partial U_t^k}}{{\partial {\alpha ^p}}}d{\alpha ^p}}, \,\,\, p \in \left\{ {1,...,P} \right\}.
\end{equation}

In the most frequently used case one has $\sigma _t^k = \sigma _t$ for any $k \in \left\{1,...,K\right\}$ and $d{\alpha ^p} = d\alpha$ for any $p \in \left\{1,...,P\right\}$ so that the formula \eqref{eq_08} can be rewritten as:

\begin{equation*}
\alpha _{t + 1}^p = \alpha _t^p + \frac{{{c_t}}}{{\sigma _t}}\delta {J_t}\sum\limits_{k = 1}^K {\delta u_t^k\frac{{\partial U_t^k}}{{\partial {\alpha ^p}}}}, \,\,\, p \in \left\{1,...,P\right\},
\end{equation*}

where $c_t = d\alpha {\gamma _t}$ is, so-called, learning rate. 

Now suppose that ${\bf{U}}\left( {{{\bf{I}}_{\tau \le t}},{{\bf{u}}_{\tau \le t}},{J_{\tau  < t}};\bsa} \right)$ is taken in the form of DNN with trainable parameters $\bsa$. In this case formula \eqref{eq_08} factually represents the analogue of well-known backpropagation algorithm for training of DNN where instead of metric gradient its estimation via random perturbations is taken. The DNN gradients $\left\{ {\partial U_t^k/\partial {\alpha ^p}} \right\}_{k = 1,p = 1}^{K,P}$ can be computed directly using any  symbolic derivative framework for neural networks (f.e., Tensorflow \cite{TF}).  

Formula \eqref{eq_08} represents the training rule for DNN ${{\bf{U}}_t}\left( \bsa \right)$ and at the same time provides SPGD-type optimization of the performance metric ${J_t}\left(\bsa \right)$. In the next sections we will give important implementation details of DNN training using formula \eqref{eq_08} as well as conduct numerical analysis of the proposed training mechanism. 

\section{Implementation details.}
\label{s2}

Application of the proposed above methodology to power beaming problem requires specifying of the mathematical formulations. For a future considerations let us suppose that the vector ${\bf{I}}_t$ of target-plane field characteristics consists of only target-plane intensity distribution so that $M=1$ and ${\bf{I}}_t = {\bf{I}}_t\left(\bsr; {{\bf{u}}} \right) = \left\{ I\left({\bsr,z = L,t} \right)\right\}$. In addition, we will suppose that the metric values are normalized to the unity interval so that ${J_t} \in \left[ {0,1} \right]$. 

\subsection{The SPGD controller.}
\label{s2_1} 

Practical implementation of SPGD algorithm based on formula \eqref{eq_06} shows that choosing of appropriate perturbation strength (variances  $\sigma_t^k$, $k \in \left\{ 1,...,K \right\}$) and gain coefficient $\gamma_t$ are extremely important for the algorithm convergence. The perturbation strength is typically chosen so that the relation $\delta J_t \sim 0.01\,\left(1 - {J_t}\right)$ for metric disturbance is held. This relation can be usually achieved using the following parametric representation for perturbation deviation $\sigma _t^k = a_\sigma ^k{\left( {1 - {J_t}} \right)^\mu } + b_\sigma ^k$, $k \in \left\{ 1,...,K \right\}$, where $a_\sigma^k \ge 0$, $b_\sigma ^k \ge 0$  and  $\mu \ge 0$ some parameters to adjust. The gain coefficient is commonly taken in the form  $\gamma_t^k = a_\gamma^k\left( {1 - {J_t}} \right) + b_\gamma^k$, $k \in \left\{ 1,...,K \right\}$, where $a_\gamma^k,b_\gamma^k \ge 0$ are also parameters for adjustment. In the most frequently used case one has $\sigma_t^k = \sigma_t$, $\gamma_t^k = \gamma_t$ for all $k \in \left\{ 1,...,K \right\}$ and the SPGD algorithm can be formulated as follows:

\begin{figure}[hbtp]
\centering
\includegraphics[width=0.5\textwidth]{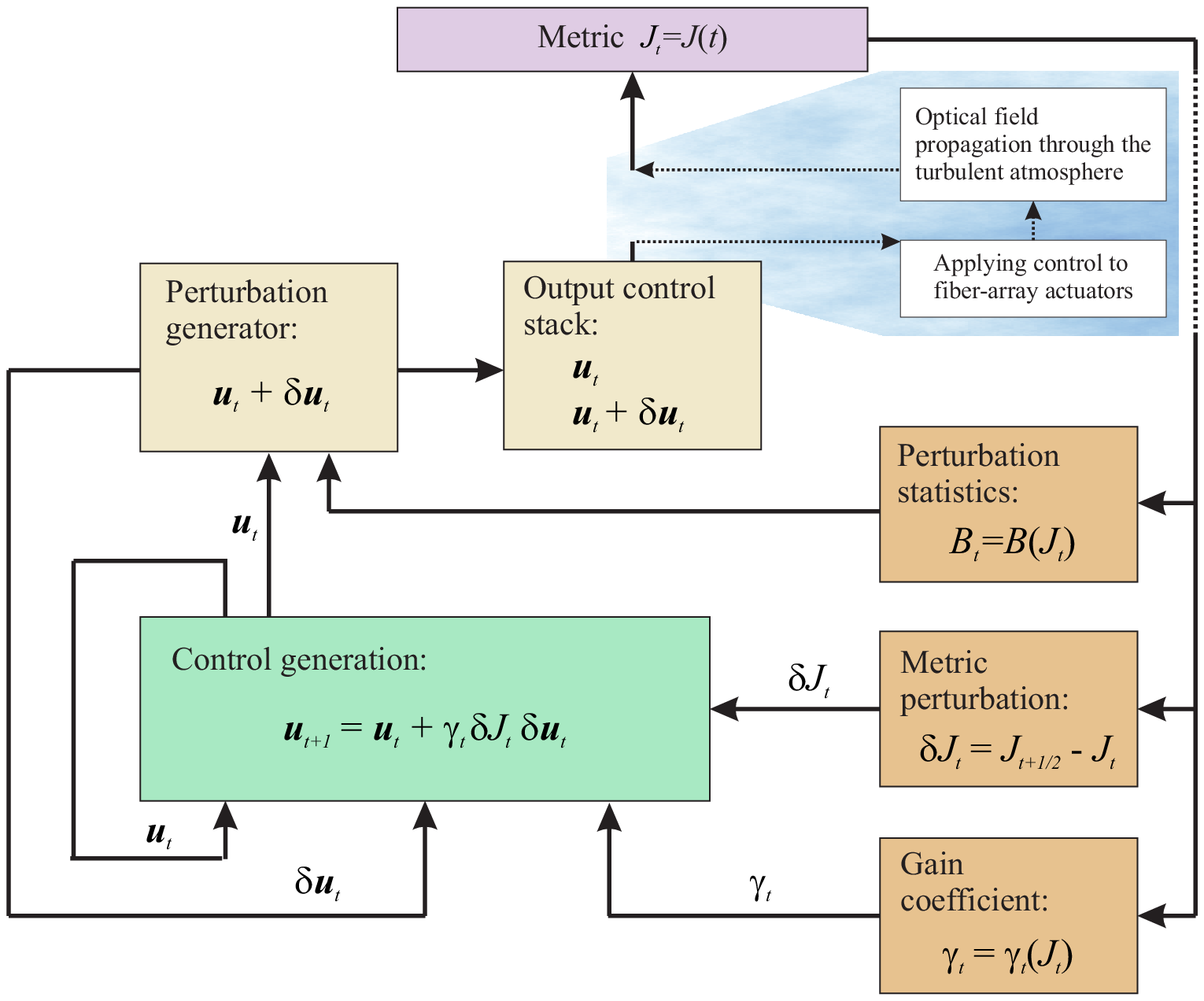}
\caption{The schematic representation of SPGD controller operational scenario.} \label{fig_03}
\end{figure}

\begin{enumerate}
\renewcommand*\labelenumi{(\theenumi)}
\setcounter{enumi}{-1}
  \item Initialization: $t=0$, ${\bf{u}}_0=0$;
	\item Either measure $J_t = J\left( {I_t} \right)$ or measure $I_t = I\left( \bsr, t;{\bf{u}}_t \right)$ and compute $J_t$;
	\item Draw random, uniformly distributed, zero-centered, vector $\delta {\bf{u}}_t$ with uncorrelated components having variance ${\sigma _t}$ each; 
	\item Either measure ${J_{t + 1/2}} = J\left( {{I_{t + 1/2}}} \right)$ or measure $I_{t + 1/2} = I\left( \bsr,t + 1/2;{{{\bf{u}}_t} + \delta {{\bf{u}}_t}} \right)$ and compute ${J_{t + 1/2}}$;
	\item Compute metric increment $\delta J_t = J_{t + 1/2} - J_t$ and $\gamma_t = a_\gamma\left( {1 - {J_t}} \right) + b_\gamma$;
	\item Update control parameters ${\bf{u}}_{t + 1} = {\bf{u}}_t + \gamma_t\delta J_t\delta {\bf{u}}_t$.
\end{enumerate}

This is conventional two-step (see explanations below) SPGD algorithm where, in order to decouple perturbation and gain factors, the multiplicative term ${\left( {\sigma _t} \right)^{-1}}$ in the formula \eqref{eq_06} is added to the gain coefficient. The operational scenario of SPGD controller is schematically shown in the Fig. \ref{fig_03}.

\subsection{The AI controller - DNN topology.}
\label{s2_2} 

\begin{figure}[hbtp]
\centering
\includegraphics[width=1.0\textwidth]{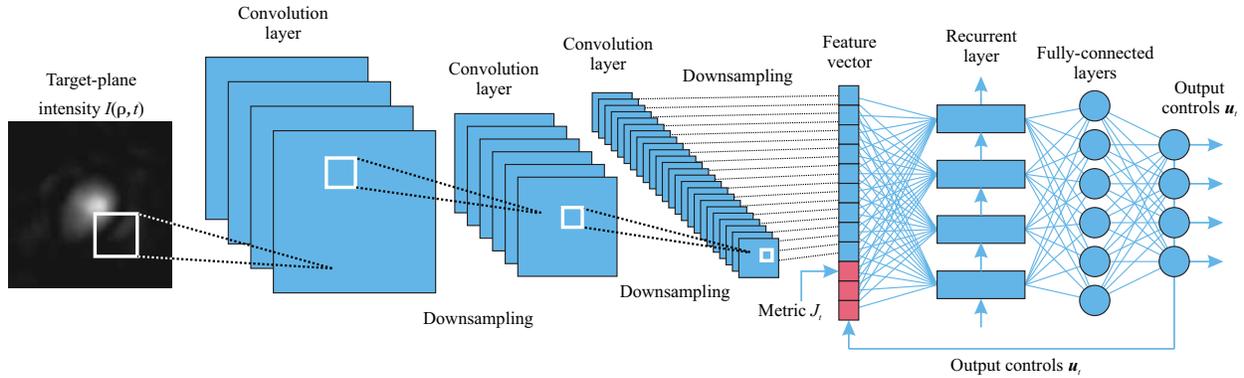}
\caption{The schematic representation of DNN topology for AI controller.} \label{fig_04}
\end{figure}

As it is usual for machine learning applications, design of AI controller begins from the specification of DNN topology that will represent the vector parametric function ${{\bf{U}}_t}\left( \bsa \right)$. Recall that AI controller for the power beaming problem will be continuously provided by measurements from PVA panel having the form of two-dimensional greyscale square image map $I_t = I\left(\bsr,t;{{\bf{u}}_t} \right)$. Supposing that this intensity distribution may contain some information useful for generation of optimal control the input DNN layers should be capable to extract this information from the image. It is well-known that the basic network structure which can provide such kind of analysis is convolutional neural network (CNN) \cite{CNN}. The CNN can be interpreted as a set of trainable digital filters dedicated to extract features from the incoming images and structurally consists of several convolution and downsampling (or pooling) layers. As far as we deal with time-dependent image flow $I_t$ the feature vector outcoming from the CNN will also keep dependency over time so that for the next and core DNN structure it is reasonable to take recurrent neural network (RNN) \cite{RNN}, especially designed for temporal data processing. The recurrent layers allow to include into analysis short- and long-term temporal relations in incoming data as well as synthesize complex and temporally correlated high-order feature vector. The conversion of feature vector outcoming from the RNN into controls is traditionally performed using several time-distributed fully-connected layers providing final analysis occurring in high-order feature space. In accordance from this design for DNN of AI controller one can propose the following topology represented in the Fig. \ref{fig_04}.

The DNN in the Fig. \ref{fig_04} has an input having the form of 4-dimensional matrix (also called as tensor in machine learning applications) of the shape $\left( {1, N_{\rm{ws}}, N_x, N_y} \right)$, where the first, so-called, batch dimension equals to 1 due to controller's operation in real-time regime, $N_{\rm{ws}} \ge 1$ is the time dimension, especially introduced for efficient training of recurrent layers, and the last two feature dimensions equal to the image size $\left( N_x, N_y \right)$. Later on, the DNN input is directed into CNN having three sequentially placed time-distributed convolutional and max-pooling layers. The CNN output, in turn, is converted into time-distributed flat feature vector. This vector is supplemented with metric value $J_t$ and DNN control outputs at the previous time step ${\bf{u}}_{t - 1}$ and, after that, is directed into stateful gated recurrent unit (GRU) \cite{GRU} containing $10K$ recurrent neurons. The GRU output is fully connected to two sequentially placed time-distributed dense (perceptron) layers with ``\textit{tanh}''-type activation function and $6K$ neurons for the internal layers and ``\textit{linear}'' activation function and $K$ neurons for the third dense output layer. It is easy to see that DNN output will have appropriate shape $\left( 1, N_{\rm{ws}}, K \right)$ for generation of $K$-dimensional control vector. Note that DNN of such particular architecture for $K=3 \cdot 19=57$ (fiber array with $N_{\rm{sa}}=19$ subapertures and piston/tip-tilts control) and $N_x=N_y= 256$ (PVA with $256\times256$ resolution) has approximately $P \sim 6\times 10^4$ trainable parameters $\bsa$.

\subsection{The AI controller -- training and inference.}
\label{s2_3} 

After choosing of DNN topology for the AI controller it is necessary to specify its training and inference algorithm. Here we present two variants of this algorithm differ from each other by SPGD approximation of metric gradient -- first algorithm (two-step algorithm) uses conventional perturbation strategy as well as the second one fuses perturbations with control trajectory (one-step algorithm). 

The two-step algorithm for inference and training of the AI controller can be formulated as follows:

\begin{enumerate}
\renewcommand*\labelenumi{(\theenumi)}
\setcounter{enumi}{-1}
\item	Initialize $t=0$, ${\bf{u}}_0=0$ and DNN trainable parameters ${\bsa}_0$;
\item Measure $I_t = I\left(\bsr, t;{\bf{u}}_t\right)$ and compute $J_t = J\left( {I_t} \right)$;
\item \textit{Training phase (optional)}:
\begin{enumerate}\renewcommand{\theenumii}{\theenumi.\arabic{enumii}}
\item Draw random, uniformly distributed, zero-centered, vector $\delta {\bf{u}}_t$ with uncorrelated components having variance ${\sigma _t}$ each;
\item Measure $I_{t + 1/2} = I\left( \bsr,t + 1/2; {{{\bf{u}}_t} + \delta {{\bf{u}}_t}} \right)$ and compute ${J_{t + 1/2}} = J\left( {{I_{t + 1/2}}} \right)$;
\item Compute metric increment $\delta J_t = J_{t + 1/2} - J_t$ and $\gamma_t = a_\gamma\left( {1 - {J_t}} \right) + b_\gamma$;
\item Update trainable parameters: \\
\begin{equation} \label{ts}
\begin{gathered}
\alpha _{t + 1}^p = \alpha_t^p + \frac{{{\gamma _t}}}{{a_\sigma}}\delta {J_t}\sum\limits_{k = 1}^K {\delta u_t^k\frac{{\partial U_{}^k\left( {{I_t},{{\bf{u}}_t},{J_t};{\bsa_t}} \right)}}{{\partial {\alpha ^p}}}}, \,\,\, p \in \left\{ {1,...,P} \right\};
\end{gathered}
\end{equation}
\end{enumerate}
\item \textit{Inference}: compute ${\bf{u}}_{t + 1} = {\bf{U}}\left( {{I_t},{{\bf{u}}_t},{J_t};{\tilde{\bsa}_{t + 1}}} \right)$, where $\tilde{\bsa}_{t + 1}={\bsa}_{t + 1}$ if training step was performed or $\tilde{\bsa}_{t + 1}={\bsa}_{t}$ if not.
\end{enumerate}

The corresponding one-step algorithm can be written as:

\begin{enumerate}
\renewcommand*\labelenumi{(\theenumi)}
\setcounter{enumi}{-1}
\item Initialize $t=0$, ${\bf{u}}_0=0$, $I_0=0$, $J_0=0$ and trainable parameters ${\bsa}_0$;
\item \textit{Inference}: compute ${\bf{u}}_{t + 1} = {\bf{U}}\left( {{I_t},{{\bf{u}}_t},{J_t};{\bsa_{t}}} \right)$;
\item \textit{Training phase (optional)}: draw random, uniformly distributed, zero-centered, vector $\delta {\bf{u}}_{t+1}$ with uncorrelated components having variance ${\sigma_{t+1}}$ each;
\item Set ${\bf{w}}_{t + 1} = {\bf{u}}_{t + 1} + \delta \tilde{\bf{u}}_{t + 1}$, where 
$\delta \tilde{\bf{u}}_{t + 1}=\delta {\bf{u}}_{t + 1}$ if training phase is turned on or $\delta \tilde{\bf{u}}_{t + 1}=0$ if not. 

\item Measure ${I_{t + 1}} = I\left( \bsr,t + 1; {{{\bf{w}}_{t + 1}}} \right)$ and compute $J_{t + 1} = J\left( I_{t + 1} \right)$;
\item \textit{Training phase (optional)}:
\begin{enumerate}\renewcommand{\theenumii}{\theenumi.\arabic{enumii}}
\item Compute metric increment $\delta J_t = J_{t + 1} - J_t$ and $\gamma_{t + 1} = a_\gamma\left( {1 - {J_{t + 1}}} \right) + b_\gamma$;
\item Update trainable parameters: \\
\begin{equation} \label{os}
\begin{gathered}
\alpha _{t + 1}^p = \alpha _t^p + \frac{{{\gamma _{t + 1}}}}{{a_\sigma ^{}}}\delta {J_{t}}\sum\limits_{k = 1}^K {\left( {w_{t + 1}^k - u_t^k} \right)\frac{{\partial U^k\left( {{I_{t + 1}},{{\bf{w}}_{t + 1}},{J_{t + 1}};{\bsa_t}} \right)}}{{\partial {\alpha ^p}}}}, \,\,\, p \in \left\{ {1,...,P} \right\}.
\end{gathered}
\end{equation}
\end{enumerate} 
\end{enumerate}

Here, as in section \ref{s2_1}, we made several simplifications with coefficients and gains. 

It is clear that one-step algorithm supposes to pass controls through the media just once in opposite to two-step version where it should be done twice per iteration. Implementation shows that in rapidly changing media this advantage can be avoid by instability of one-step iteration process and corresponding control trajectories.

\begin{figure}[hbtp]
\centering
\includegraphics[width=0.6\textwidth]{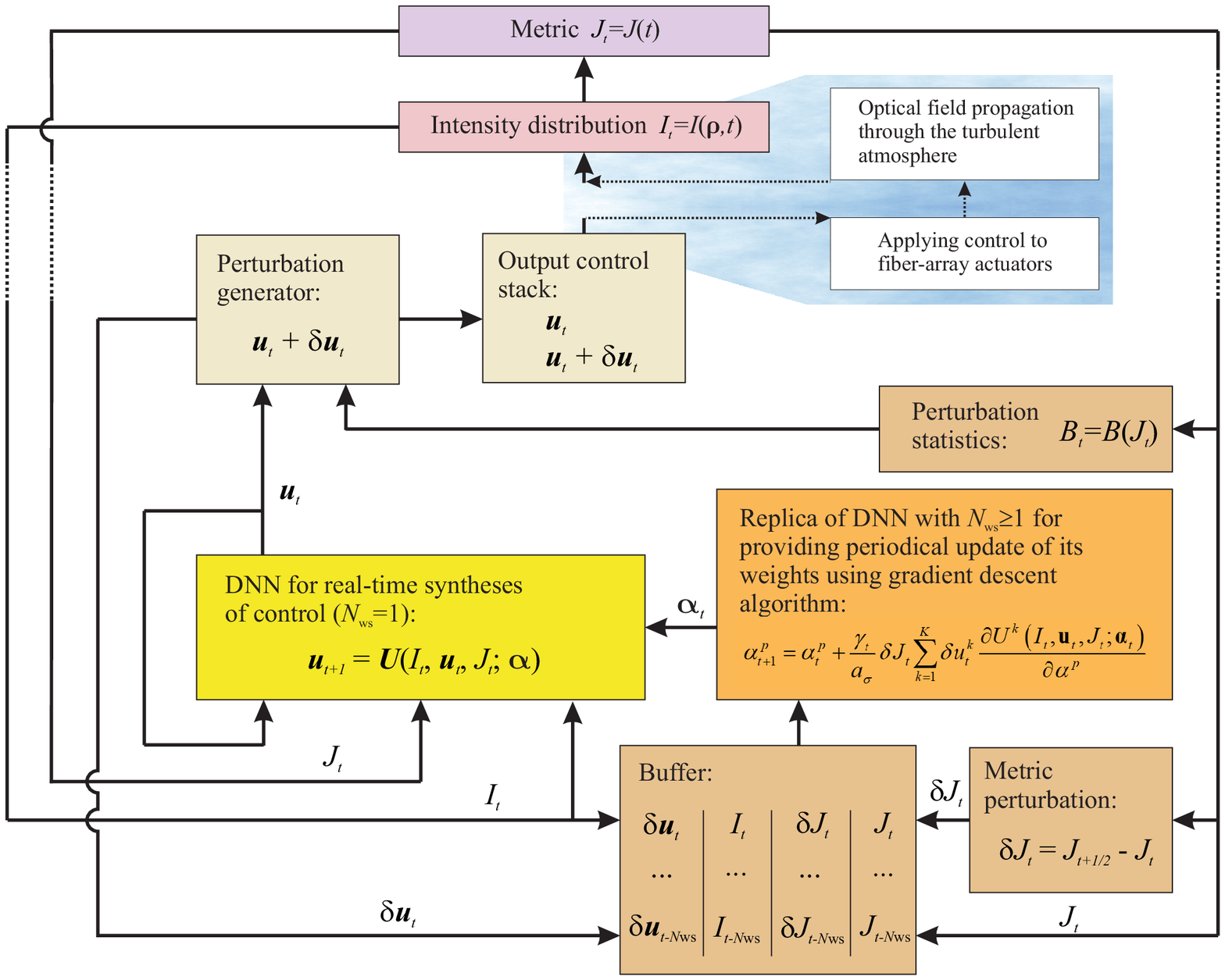}
\caption{The schematic representation of AI controller operating in training mode.} \label{fig_05}
\end{figure}

The proposed algorithms structurally combine two regimes: DNN inference for synthesizing optimal control as well as DNN weights updating for training. If the controller operates in pure inference mode, when DNN weights updating is off, the controller uses previously collected knowledge to control the system. In training mode control synthesis as before is provided via DNN inference however DNN weights updating mechanism additionally provides SPGD-type optimization occurring now in training parameters $\bsa$ space  instead of initial control parameters ${\bf{u}}$ space as in conventional SPGD algorithm. As we will see below this combination of optimization and training in one process leads to a quite interesting numerical effects.  

Utilizing of RNN layers in DNN topology as well as providing stability of training process require to have two replicas of DNN (represented as computational graphs) sharing one set of trainable parameters: one copy for inference mode, where the optimal control is synthesized, with $N_{\rm{ws}}=1$ and another copy for training mode with $N_{\rm{ws}}\ge 1$. Note that, in case $N_{\rm{ws}} > 1$ weights updating increment in formulas \eqref{ts} and \eqref{os} should be slightly modified by adding extra summation over sequential $N_{\rm{ws}}$ time steps. Updating of the training parameters ${\bsa_{t}}$ in both algorithms can be occurred by any reasonable strategy (periodically, by achieving some metric value and so on) and supposes to use a buffer of the size $N_{\rm{ws}}$ containing all DNN inputs as well as estimations of metric gradient with the corresponding perturbations. 

The detailed illustration of AI controller operational scenario is given in the Fig. \ref{fig_05}. 

\subsection{Combined SPGD and AI control.}
\label{s2_4}

In the consideration above it was implicitly supposed that the controller has all necessary information for synthesis of correct control in any time. However, real-world physical systems, as a rule, operate in conditions when information available for the analysis is not full and the control based on this information can be non-unique. In addition, realistic systems usually operate in presence of stochastic and unpredictable factors that can not be accounted using deterministic control generators. In this case the proposed approach has to be supplemented by adding some mechanism that can compensate such unpredictable stochastic factors as well as the lack in input information.     

Consider slightly modified formula for the control in the form:

\begin{equation*}
 {\bf{u}}_t = {\bf{w}} + {\bf{U}}_t \left(\bsa, {\bf{w}} \right),
\end{equation*}

where ${\bf{w}}=\left(w^1,...,w^K \right)$ is the part of the control responsible for compensation of aforementioned factors. Then, the optimization problem \eqref{eq_07} can be rewritten as: 

\begin{equation*}
\begin{gathered}
J_t\left( {\bf{w}}, \bsa \right) = J\left( {\bf{I}}_t \left( \bsr; {\bf{w}}+ {\bf{U}}_t \left(\bsa, {\bf{w}} \right) \right) \right),\\
J_t \left( {\bf{w}}, \bsa \right) \to \max\limits_{\bsa, {\bf{w}}}. 
\end{gathered}
\end{equation*}

Under condition of uniform boundedness of DNN derivative over ${\bf{w}}$ variables, i.e. $\left| \partial U^k/\partial w^l \right | \le C$, for some constant $C>0$ and any $k, l\in\left\{ 1,...,K \right\}$, the optimal control trajectory can be approximated as:

\begin{equation*}
\begin{gathered}
{\bf{w}}_{t + 1} = {\bf{w}}_{t} + \frac{{{\gamma _t}}}{{\sigma_t}}\delta {J_t}\delta{\bf{u}}_t, \\
{\bf{v}}_{t + 1}= {\bf{U}}\left( {{I_t},{{\bf{v}}_{t}}, {{\bf{w}}_{t + 1}},{J_t};{\bsa_{t}}} \right), \\
{\bf{u}}_{t + 1}  = {\bf{w}}_{t + 1}  + {\bf{v}}_{t + 1},
\end{gathered}
\end{equation*}

so that we factually have combination of DNN-based and SPGD control. The corresponding training and inference algorithm in this case can be written as:

\begin{enumerate}
\renewcommand*\labelenumi{(\theenumi)}
\setcounter{enumi}{-1}
\item	Initialize $t=0$, ${\bf{w}}_0=0$, ${\bf{v}}_0=0$ and DNN trainable parameters ${\bsa}_0$;
\item Calculate ${\bf{u}}_t={\bf{w}}_t + {\bf{v}}_t$;
\item Draw random, uniformly distributed, zero-centered, vector $\delta {\bf{u}}_t$ with uncorrelated components having variance ${\sigma _t}$ each;
\item Measure $I_{t + 1/2} = I\left( \bsr,t + 1/2; {{{\bf{u}}_t} + \delta {{\bf{u}}_t}} \right)$ and compute ${J_{t + 1/2}} = J\left( {{I_{t + 1/2}}} \right)$;
\item Compute metric increment $\delta J_t = J_{t + 1/2} - J_t$ and $\gamma_{t} = a_\gamma\left( {1 - {J_{t}}} \right) + b_\gamma$;
\item \textit{SPGD optimization}: calculate 
\begin{equation*}
\begin{gathered}
{\bf{w}}_{t + 1} = {\bf{w}}_{t} + \frac{{{\gamma _t}}}{{\sigma_t}}\delta {J_t}\delta{\bf{u}}_t;
\end{gathered}
\end{equation*}
\item \textit{Training phase(optional)}: update trainable parameters \\
\begin{equation*}
\begin{gathered}
\alpha _{t + 1}^p = \alpha_t^p + \frac{{{\gamma _t}}}{{a_\sigma}}\delta {J_t}\sum\limits_{k = 1}^K {\delta u_t^k\frac{{\partial U_{}^k\left( {{I_t},{{\bf{v}}_t}, {{\bf{w}}_{t+1}},{J_t};{\bsa_t}} \right)}}{{\partial {\alpha ^p}}}}, \,\,\, p \in \left\{ {1,...,P} \right\};
\end{gathered}
\end{equation*}
\item \textit{Inference}: compute ${\bf{v}}_{t + 1} = {\bf{U}}\left( {{I_t},{{\bf{v}}_t}, {{\bf{w}}_{t+1}}, {J_t};{\tilde\bsa_{t + 1}}} \right)$, where $\tilde{\bsa}_{t + 1}={\bsa}_{t + 1}$ if training step was performed or $\tilde{\bsa}_{t + 1}={\bsa}_{t}$ if not.
\end{enumerate}

Note that, in general case SPGD- and DNN-based controls can be multiplied by some gains in order to effectively balance a contribution of each mechanisms in optimization. 

\subsection{Regularization.}
\label{s2_5} 

It is well-known that control systems operated in accordance with active feedback control scheme have a tendency to forming of positive feedback loops that leads to significant instability in its operation and training. The AI controller is not an exception to this rule and weights updating mechanism represented above does not guarantee an absence of positive feedbacks in control. The traditional solution avoiding this problem is utilizing of regularized optimization metrics as well as smoothing of metric gradients. In order to realize these ideas, at first, let us consider two functionals. First functional is designed to bound deviations in output control trajectories during training process and has the form: 

\begin{equation}\label{S}
{S_t}\left( \bsa \right) = \gamma _{}^S\sum\limits_{k = 1}^K {{{\left( {\frac{{dU_t^k\left( \bsa \right)}}{{dt}}} \right)}^2}}, 
\end{equation}

where $\gamma^S \ge 0$ is the gain (typically, $\gamma^S \sim {10^{-2}}$). Note that in \eqref{S} exactly the full derivative of functions $U_t^k\left( \bsa \right)$, $k \in \left\{ {1,...,K} \right\}$ over time is placed in order to avoid instability originated not only from high-order oscillations of controller's output but from metric and DNN input jitters too. 

In order to additionally smooth the output control as well as to prevent DNN from the overtraining the traditional $L_2$ functional applying directly to DNN training weights is used:

\begin{equation*}
{L_2}\left( \bsa \right) = \gamma _{}^L{\left\| \bsa \right\|^2} = \gamma _{}^L\sum\limits_{p = 1}^P {\alpha _p^2},
\end{equation*}

where $\gamma^L \ge 0$ is the gain (typically, $\gamma^L \sim {10^{-3}}$).

Finally, training of DNN using formula \eqref{eq_08} can meet convergence process instability due to chaotic changing of random perturbations. In order to avoid these gradient oscillations the standard momentum \cite{Mom} or Adam \cite{Adam} algorithm can be applied. Combining gradient smoothing with metric regularization one can rewrite weights updating step as: 

\begin{enumerate}
\renewcommand*\labelenumi{(\theenumi)}
\setcounter{enumi}{-1}
\item	Initialize $t=0$, $g_0^p=0$;
\item Update trainable parameters: \\
\begin{equation*}
\begin{gathered}
g_{t + 1}^p = \nu g_t^p + \frac{\partial }{{\partial {\alpha ^p}}}\left( {\sum\limits_{k = 1}^K {\left( {\frac{{{\gamma _t}}}{{a_\sigma ^{}}}\delta {J_t}\delta u_t^kU_t^k - \gamma _{}^S{{\left( {U_t^k - U_{t - 1}^k} \right)}^2}} \right)} } \right) - 2\gamma _{}^L\sum\limits_{p = 1}^P {\alpha _p^{}};
\end{gathered}
\end{equation*}
\item Calculate  $\alpha _{t + 1}^p = \alpha _t^p + g_t^p$, $p \in \left\{ {1,...,P} \right\}$.
\end{enumerate}

Here $\nu \in \left[ {0,1} \right)$ is smoothing factor typically chosen as  $\nu = 0.9$. Note that the increment in the step (1) is specially given in the form where partial derivatives over trainable weights are taken from  the weighted sum of DNN outputs that maximally convenient for implementation using popular machine learning frameworks. 

\subsection{Decoupling of control channels.}
\label{s2_6} 

Returning back to physical statement of the problem it is necessary to note that the configuration of control parameters represented by the formula \eqref{eq01} provides extremely strong coupling between control channels. This situation is connected with physics of fiber-array beamlets propagation given by equation \eqref{Parab}. For example, effective target-plane focusing of the fiber-array at time $t \ge 0$ needs to configure pistons and tip-tilts so that the sum 
$ \sum\limits_{n=1}^{N_{\rm{sa}}} H_n\left(\bsr\right)\varphi_n \left(\bsr, t\right)$ on fiber-array aperture will approximate paraboloid function representing thin lens. The same principle is fulfilled and for the control of focal spot displacement -- in this case one need to approximate the sum of paraboloid and linear function.

It is well-known that robust synthesis of control for a system with coupled channels is much more complicated problem in comparison with control of the same system but with decoupled channels. As it was noted before modern SPDG controller provides extremely high iteration rate and mentioned problem is not quite actual. However, for AI control this problem can be actual especially taking into account that here the optimization process is performed in $N_{\rm{ws}}$ times less often regarding to SPGD. To avoid this drawback let us partially decouple fiber-array control channels introducing following decomposition of control parameters. 

At pupil-plane $z=0$ consider fiber-array aperture (see Fig. \ref{fig_02}, left) and let $\left\{ {{Z_q}\left( \bsr \right)} \right\}_{q = 1}^{Q}$ be the family of mutually-orthogonal and mean-square normalized Zernike polynomials on the aperture disc, where 
$Q={\left( N_Z + 1 \right)\left( N_Z + 2 \right)/2 - 1}$ and $N_Z\ge 1$ is maximal degree of polynomials in this set \cite{Zern}. It is well-known that first five Zernike polynomials ($N_Z=2$) represent fundamental optical aberrations including $x$- and $y$- slopes and defocus and hence can be considered as a basis for reducing complexity of the light-spot control. Extending this assumption to the polynomials of degree more than 2 let us build the transformation between coefficients in Zernike polynomials space and fiber piston/tip-tilt basis. For $q \in \left\{ {1,...,Q} \right\}$ define:  

\begin{equation*}
\begin{aligned}
r_{kq} &= \frac{4}{\pi d^2}
\begin{cases}
\int {{H_k}\left( \bsr \right){Z_q}\left( \bsr \right){d^2}\bsr}, &\,\,\, k \in \left\{ {1,...,{N_{{\rm{sa}}}}} \right\}, \\[5pt]
\int{(x-x_{k-N_{\rm{sa}}})H_{k-N_{\rm{sa}}}\left( \bsr \right){Z_q}\left( \bsr \right){d^2}\bsr}, &\,\,\, k \in \left\{ {{N_{{\rm{sa}}}} + 1,...,2{N_{{\rm{sa}}}}} \right\},\\[5pt]
\int{(y-y_{k-2N_{\rm{sa}}})H_{k - 2N_{\rm{sa}}}\left( \bsr \right){Z_q}\left( \bsr \right){d^2}\bsr}, &\,\,\,  k \in \left\{ {2{N_{{\rm{sa}}}} + 1,...,3{N_{{\rm{sa}}}}} \right\}.
\end{cases}
\end{aligned}
\end{equation*}

Then, the matrix $R = \left[ {{r_{kq}}} \right]_{k = 1,q = 1}^{K,Q}$ of the size $K \times Q$ provides transformation between Zernike control coefficients and conventional piston/tip-tilt representation so that ${\bf{u}} = {{\bf{u}}_{Z}}R$, where ${{\bf{u}}_Z}$ is $Q$-dimensional control vector in Zernike space. 

Note that, for any particular configuration of fiber-array subapertures the matrix $R$ is computed once, each transformation from Zernike representation to fiber-array controls is required just one matrix multiplication by a vector and can be performed extremely fast even comparing with SPGD iteration speed.   

\section{Numerical results.}
\label{s3}

\subsection{Numerical verification of AI controller's training capabilities.}
\label{s3_1} 

In order to verify training and optimization capabilities of the proposed AI-based control approach let us consider simple tracking system. Let $M = 1$, $K=2$ and 

\begin{equation}\label{I}
{I_t}\left( {x,y;{u^1},{u^2}} \right) = {e^{ -{\left[ {{{\left( {x - x\left( t \right) + u_{}^1} \right)}^2} + {{\left( {y - y\left( t \right) + u_{}^2} \right)}^2}} \right]/\beta_1^2}}},
\end{equation}

where $\bsr\left( t \right) = \left( {x\left( t \right),y\left( t \right)} \right)$ is some predefined trajectory in $Oxy$ coordinates and ${\beta_1} > 0$. From the formula \eqref{I} it is clear that ``intensity'' distribution simulated by the formula \eqref{I} under control ${\bf{u}}_t=\left( u^1(t), u^2(t)\right)$ will represent a small spot located at the point $\left(x(t)-u^1(t), y(t)-u^2(t)\right)$ for any $t\ge 0$. Let  $\bsr\left( t \right) = \left( {\sin \left( {\omega t} \right),\cos \left( {\omega t} \right)} \right)$ with some $\omega > 0$ so that circular motion of the spot is simulated. 

As an tracking objective consider the problem of ``catching and holding'' of the spot at the coordinates origin. In this case for the performance metric one can take:

\begin{equation}\label{M}
J\left( {{I_t}\left( x,y;{{u^1},{u^2}} \right)} \right) = \frac{1}{\pi}\left( \frac{1}{\beta_1^2} + \frac{1}{\beta_2^2}\right)\int
{{I_t}\left(x,y;{{u^1},{u^2}} \right){e^{ -\left( {{x^2} + {y^2}} \right)/{\beta_2^2}}}dxdy}, 
\end{equation}

where ${\beta_2} > 0$ defines the metric tolerance to the spot position. It is easy to see that metric \eqref{M} stimulates to synthesize control trajectory ${\bf{u}}_t$ maximally closed to spot trajectory $\bsr\left( t \right)$ and, moreover, this metric is equivalent to smooth Strehl ratio introduced before. For the AI controller consider the same DNN as has been proposed in section \ref{s2_2}. 

In the numerical experiments the simulation area was $\left[-5.0, 5.0\right]_x \times \left[-5.0, 5.0 \right]_y$ and had 256x256 pixels resolution, $\beta_1 = 0.4$ and  $\beta_2 = 1.0$, the simulation time step was 0.1 sec so that the frequency factor $\omega$ provided 1 full rotation of the spot per approximately 5 nominal minutes. The DNN grayscale input image had the same 256x256 pixel resolution, $N_{\rm{ws}} = 4$ and total number of trainable parameters for this DNN was about $4.5\times10^4$. The simulations were performed in Python v.3.7 \cite{Python}, the simulator was completely separated from the controller which was implemented using Tensorflow v.1.14 \cite{TF} library for Python. For inference and training of the AI controller the two-step algorithm represented in section \ref{s2_3} was used. Overall simulation and training time on ASUS ROG Zephyrus\footnote{Intel Core i7-9750H 2.60GHz, 32 Gb RAM with Nvidia GeForce RTX 2080 with Max-Q Design, 8Gb VRAM.} took about 30 seconds on 1 minute of simulation time. 

\begin{figure}[hbtp]
\centering
\includegraphics[width=0.8\textwidth]{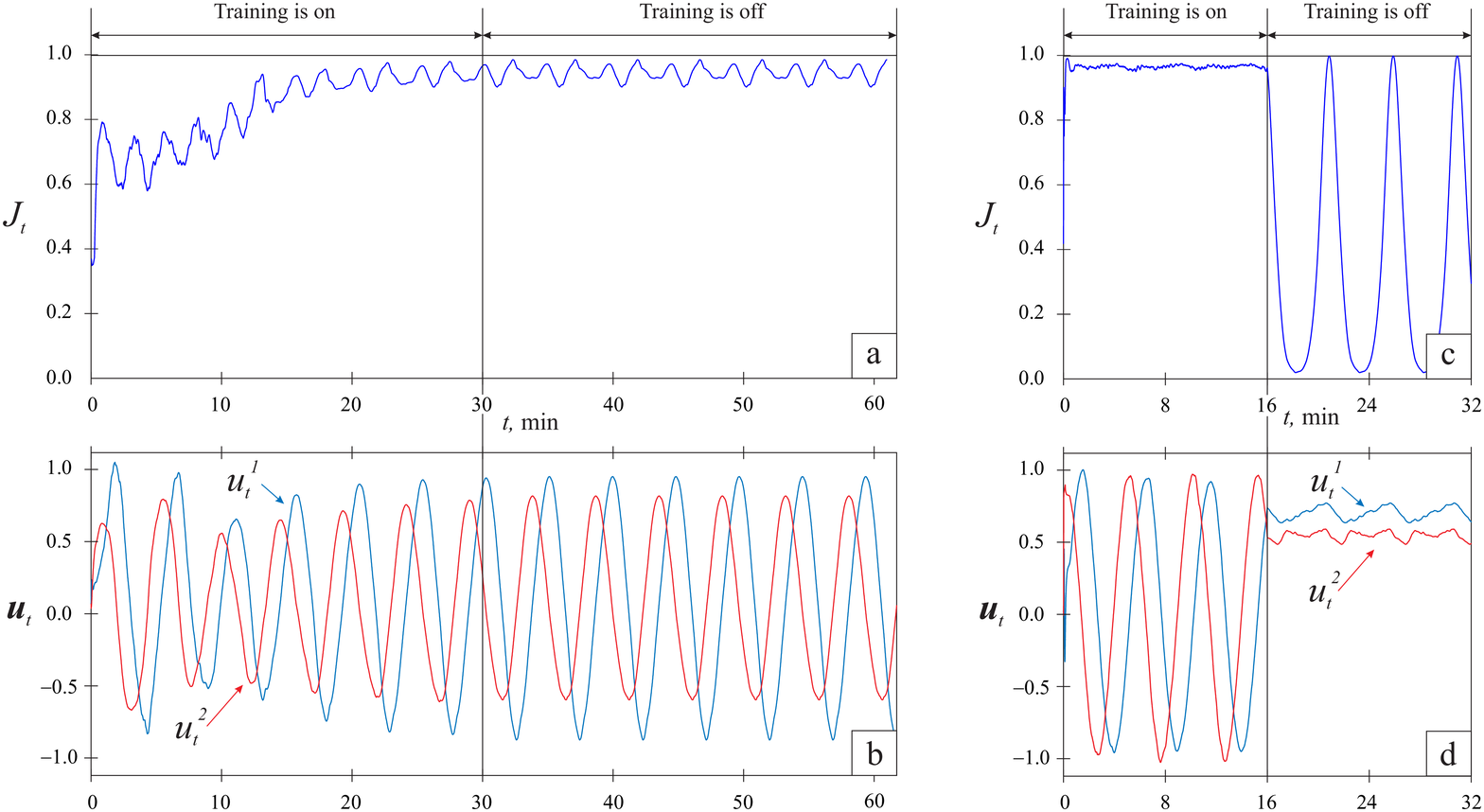}
\caption{Performance metric $J_t$ (a, c) and trajectories ${{\bf{u}}_t}$ (b, d) versus time for two learning strategies in the tracking task. Charts (a, b) represent ``soft'' training strategy with relatively small learning rate and perturbation strength. In opposite, in charts (c, d) more ``aggressive'' strategy is chosen in order to rapidly achieve metric maximum.} 
\label{fig_06}
\end{figure}

The Fig. \ref{fig_06} represents numerical results of AI control performance and corresponding control trajectories for two different optimization and learning strategies. The first strategy (Fig. \ref{fig_06}, a--b) supposes to make relatively slow motions in the spot's direction with relatively small learning rate and perturbation strength. In opposite to this scenario the second strategy (Fig. \ref{fig_06}, c--d) has the aim of maximally rapid ``catching'' of the spot and accompany it in its motion. It is easy to see that just the first strategy allows controller to learn the tracking principle and hence allows it to operate without SPGD-type iterations. The second strategy failed teaching the controller so that turning off the SPGD iterations immediately led to loosing of the spot. Moreover, continuing using the controller in such aggressive regime without significant regularization rapidly fell the controller to instability with generation of ``saw''-type control.

Note that the same results were achieved using Gaussian randomly drawn two-parametric process $\bsr\left( t \right)$ with mutually uncorrelated components\footnote{For this scenario Python code is available for testing by the request to authors.}. 

\begin{figure}[hbtp]
\centering
\includegraphics[width=0.6\textwidth]{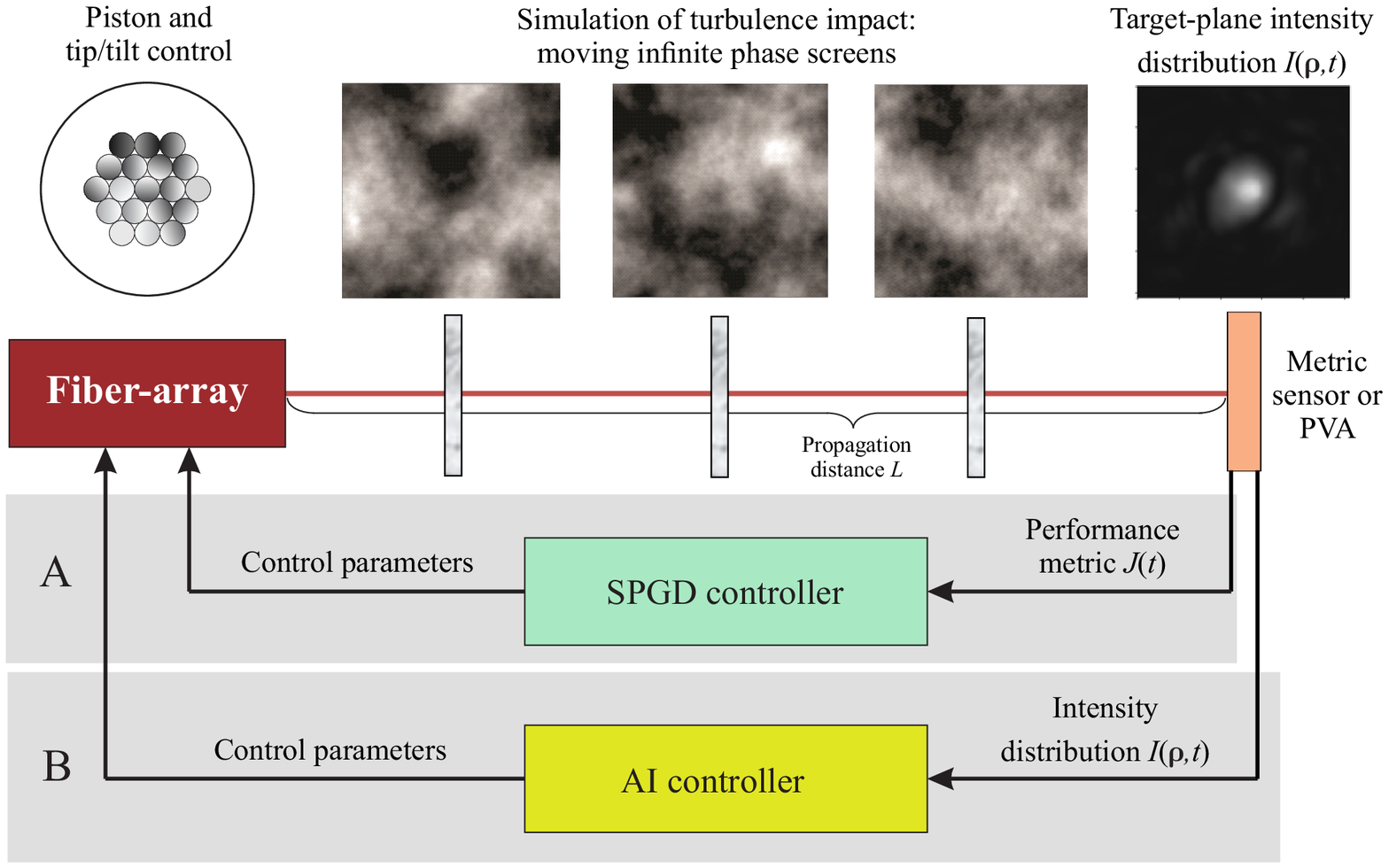}
\caption{Illustration of simulation setup for fiber-array power beaming problem in presence of atmospheric turbulence. The control of fiber-array output field can be performed either (A) SPGD control system or (B) AI controller.} 
\label{fig_07}
\end{figure}

\subsection{Numerical results for power beaming problem.}
\label{s3_2} 

Consider hexagonal fiber-array consisting of $N_{\rm{sa}} = 19$ densely packed subapertures of the diameter $d = 60$ mm each with distances between subapertures equals also to 60 mm (Fig. \ref{fig_02} on the left) and Gaussian beamlet diameters $a_0 = 0.9 d$. In our analysis, we consider horizontal propagation scenario at the distance $L = 5000$ m and vary the refractive-index-structure parameter characterizing homogeneous atmospheric turbulence strength,  $C_n^2$, in the range $5.0\times10^{-16}-1.5\times10^{-15} \,\text{m}^{-2/3}$. Simulation of atmospheric turbulence changes in time is realized by introducing of wind transversal to the propagation direction with constant speed $w$ in the range $1-6$ m/s. The Airy diameter at the target-plane is $d_{\rm{a}}=40$ mm for wavelength $\lambda=1.064\, \mu\text{m}$  ($k=5.905 \times 10^6\, \text{m}^{-1}$) and target-plane PVA has square shape with the side $D=200$ mm.

The numerical simulations of atmospheric propagation of the optical wave from fiber-array to PVA is performed through numerical integration of \eqref{Parab} using so-called split-operator method \cite{FlMoFe, Sch} when the impact of turbulence-induced aberrations is modeled with equidistantly placed thin phase screens (see Fig. \ref{fig_07}). The conventional approach \cite{FlMoFe} assumes using of phase screens with finite size in both $x$ and $y$ directions. The simulation of atmospheric wind requires continuously shifting of these screens in $Oxy$ plane with the need of their extension if wind-induced displacement will exceed of the screen size. This extension traditionally can be done either using periodical (in both directions) finite phase screens or utilizing phase screens infinitely-long in one specified direction \cite{VoParVV}. In our simulations we use both approaches considering propagation scenario with periodic phase screens as a reduced complexity control task. The number of phase screens in numerical simulations is varied in the range 5-10.

\begin{figure}[hbtp]
\centering
\includegraphics[width=0.8\textwidth]{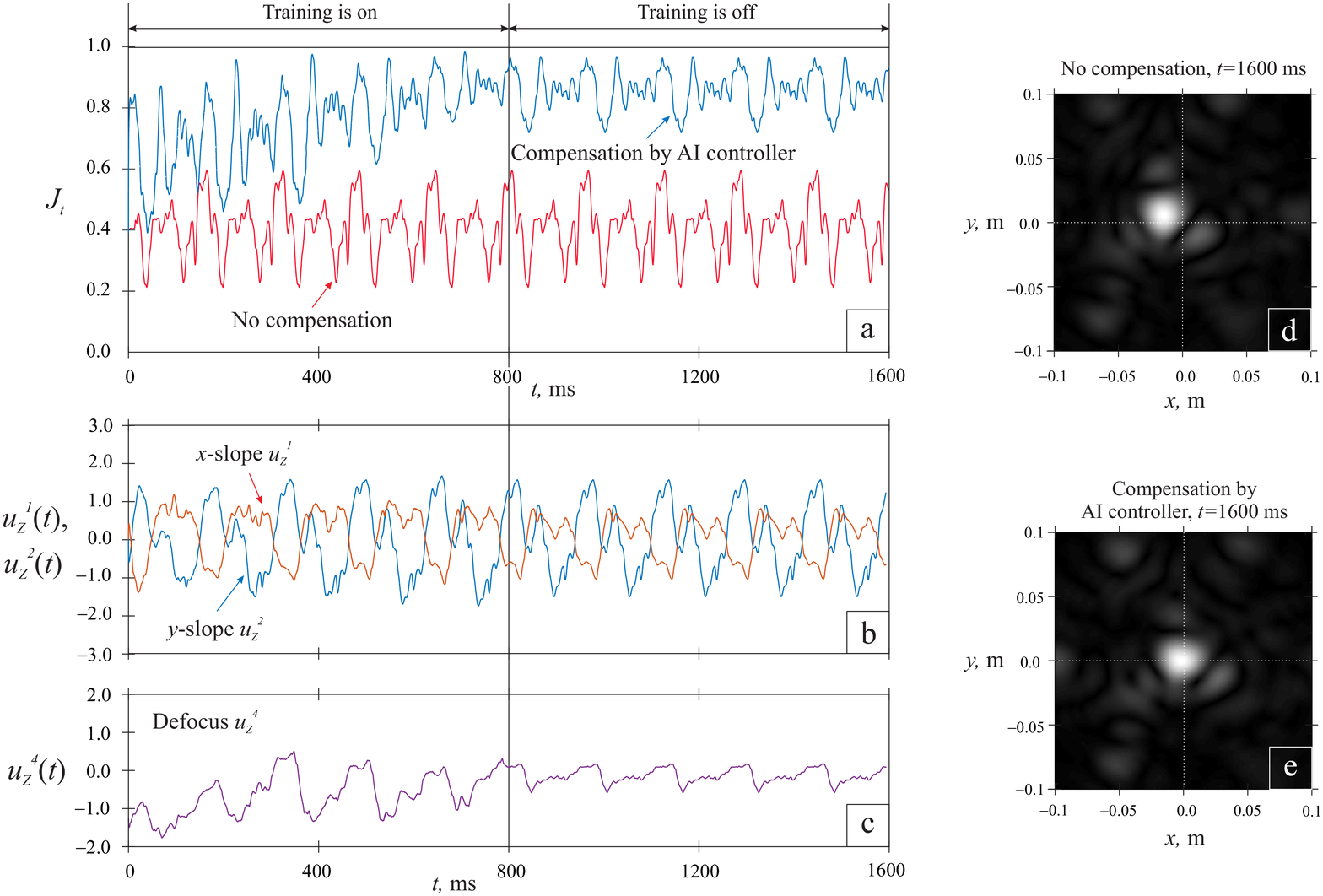}
\caption{The chart (a) represents comparison of AI controller compensation performance with ``no compensation'' curve taken for the period $[0,1600]$ ms. The AI controller operates in both training (during first 800 ms) and inference (during the following 800 ms) modes, control is performed in Zernike space using dimensionless coefficients at first $Q=14$ Zernike polynomials and periodical phase screens were taken for simulation of atmospheric turbulence. The charts (b) and (c) represent trajectories for $x$- and $y$- slopes ($u_Z^1$ and $u_Z^2$ control variables) and defocus ($u_Z^4$ variable) versus time. The images (d) and (e) represent square roots from intensity distributions taken in PVA area $[-0.1, 0.1]_x \times [-0.1, 0.1]_y$ $\text{m}^2$ at $t=1600$ ms for system operating with  no compensation (d) and with AI controller (e). In this numerical experiment $C_n^2=1.0\times10^{-15}, \,\text{m}^{-2/3}$ and wind speed $w=5$ m/s.} 
\label{fig_08}
\end{figure}

The simulation square in $xy$-plane is centered at the origin, completely covers PVA area and has 800 mm in both directions with corresponding numerical grid having 1024x1024 pixels resolution. The phase screen resolution is chosen the same. The simulation time step sets to $5.0\times10^{-5}$ sec so that the SPGD controller is factually operated at $2\times10^4$ iteration per second and DNN training has  $2\times10^4/N_{\rm{ws}}$ weight updates per second. The DNN input is provided by PVA images that have 256x256 pixels resolution. The window size for the training DNN's replica is set to $N_{\rm{ws}}=4$.

The performance metric is chosen as smooth Strehl ratio introduced in section \ref{s1_2}. The numerical integration in the formulas \eqref{J}-\eqref{Jvac} is performed over PVA area only. The metric smoothing factor is $\beta={d_{\rm{a}}}/4$.

The fiber-array is supplied by a optimization controller capable to operate with phase pistons and tip-tilts, so that overall number of control parameters is $K = 3 \times N_{\rm{sa}} = 57$. In addition, we will assume that this controller is capable to convert controls from Zernike space to piston/tip-tilt representation, i.e. compute the transformation ${\bf{u}} = {{\bf{u}}_{Z}}R$ (see section \ref{s2_6}), and this transformation can be done with no additional time consumption. In our numerical experiments we utilize the family of Zernike polynomials with degree $N_Z\le 4$ ($Q=14$ polynomials totally) including: $x$- and $y$- slopes (coefficients $u_Z^1$ and $u_Z^2$, correspondingly), oblique astigmatism ($u_Z^3$), defocus ($u_Z^4$) and vertical astigmatism ($u_Z^5$). Note that numerical simulation and modeling of fiber-array optical system was performed using WONAT software library \cite{Wonat} adapted for Python 3.7.

In our analysis we consider four scenarios of control system operation: (1) propagation with no compensation, (2) compensation by SPGD algorithm (Fig. \ref{fig_07}, A), (3) compensation by AI controller in training mode and (4) compensation by AI controller purely in inference mode (Fig. \ref{fig_07}, B for both modes). 

\begin{figure}[hbtp]
\centering
\includegraphics[width=0.6\textwidth]{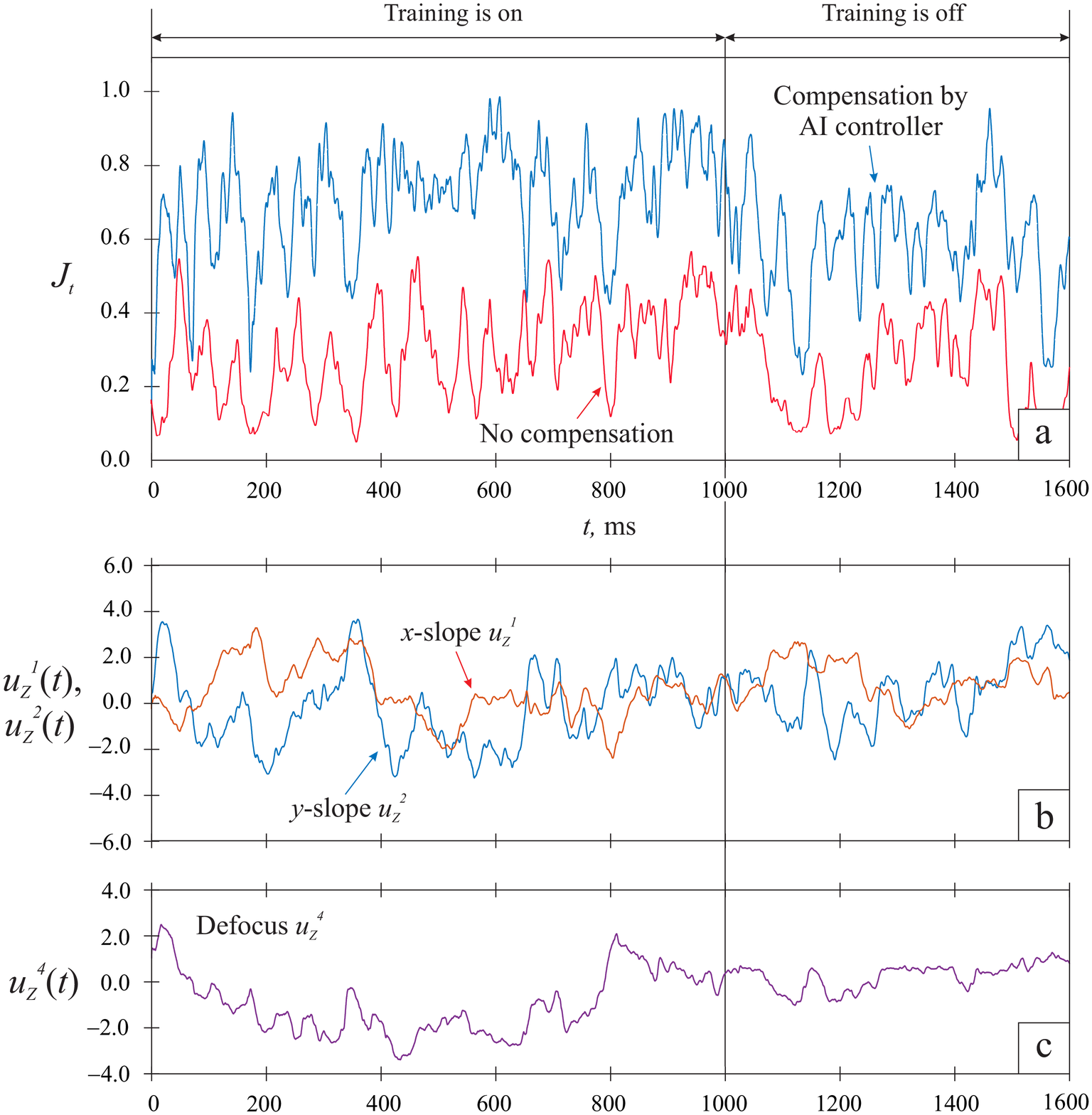}
\caption{The chart (a) represents comparison of AI controller compensation performance with ``no compensation'' curve taken for the period $[0,1600]$ ms. Atmospheric turbulence is modeled using infinite aperiodic phase screens and control is performed in Zernike space with $Q=14$ polynomials. Training phase of AI controller continues for a first simulation second and all remaining time controller operates in inference mode. The charts (b) and (c) represent trajectories for $x$- and $y$-slopes ($u_Z^1$ and $u_Z^2$ control variable) and defocus ($u_Z^4$ variable) versus time. Here $C_n^2=1.0\times10^{-15}, \,\text{m}^{-2/3}$ and wind speed $w=5$ m/s.} 
\label{fig_09}
\end{figure} 

In order to be assure that AI controller is configured properly and its operating is accompanied by its learning, at first, consider 5 finite and periodic phase screens slowly moving by constant wind $w = 5$ m/s. The total observation period is taken as 1600 ms so that phase screens have a time to make 10 complete cycles. Overall time interval is separated into two equal sub-intervals -- for controller training and its inference with no training. The Fig. \ref{fig_08} represents the results of this experiment. During first 800 ms in the training mode the controller reaches approximately $\left\langle J_t\right\rangle=0.8$ compensation performance via active control of Zernike coefficients including $x$- and $y$- slopes and defocus. Turning the training off after 800 ms keeps compensation approximately at the same level, however, the most important that control variables are continuing to repeat the control patterns learned during training period. This example demonstrates that even for such a short training time the controller was able to remember one of the possible optimization strategy completely rely on observation of target-plane image distributions.

The next numerical experiment is dedicated to verify the controller's generalization properties -- can the controller learn the control strategy in conditions of aperiodic changing of atmospheric turbulence. For this task let us consider 6 infinite aperiodic phase screens moving by constant wind $w=6$ m/s and compare compensation results of AI controller with ``no compensation'' metric curve. The Fig. \ref{fig_09} represents results of this experiment. During the first simulation second the performance of AI control oscillates approximately around the value $\left\langle J_t\right\rangle=0.7$ and a wide spread observed for both of curves can be explained by using of infinitely-long phase screens having large turbulence outer-scale \cite{VoParVV}. During this period the controller is actively looking for optimal trajectories of control variables that follows from the form of curves represented in Fig. \ref{fig_09}, (b) and (c) on the left of solid vertical line. After turning the training off the overall compensation level is degraded to $\left\langle J_t\right\rangle=0.5$ however the controller proceeds active changing of phase slopes in order to place target-plane light spot in central position. In parallel with slopes correction the controller is tried to optimize the spot size by control of beam focusing however is doing that not so active as before either due to sufficient focusing during training phase or poor training of this capability. Thus, this example explicitly shows that the controller is capable to rather prompt ``understand'' elementary relations between phase slope and target spot position, beam focusing and spot size and can account these relation during synthesis of the control in inference mode.

\begin{figure}[hbtp]
\centering
\includegraphics[width=0.6\textwidth]{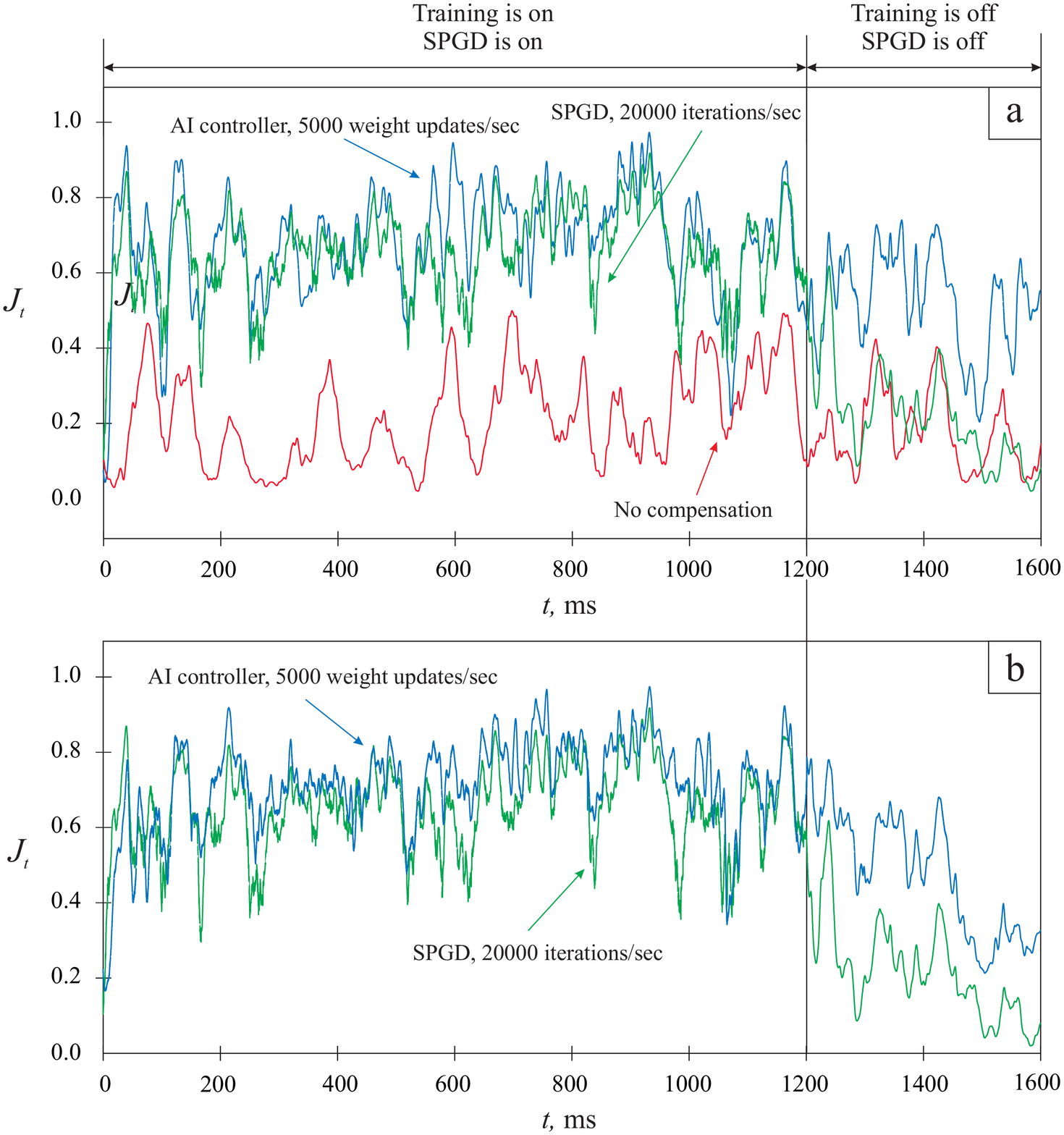}
\caption{The chart (a) represents comparison of SPGD- and AI-based control system performance, where infinite aperiodic phase screens were used and both algorithms are operated with piston and tip-tilt control. The SPGD optimization as well as training phase of AI controller continue for a first 1200 ms and all remaining time SPGD is turned off and AI controller operates in inference mode. The chart (b) represents analogous metric comparison with the same phase screens realizations as in (a) but for more ``aggressive'' training strategy of AI controller. Here $C_n^2=1.0\times10^{-15}, \,\text{m}^{-2/3}$ and wind speed $w=4$ m/s.} 
\label{fig_10}
\end{figure}

The comparison of AI and SPGD control performance will be done in two steps. Let the control is performed directly using piston/tip-tilt variables. In the first experiment we will mostly focus on AI controller learning capabilities instead of its compensation performance and consider a ``soft'' training strategy with constant learning rate $\gamma_t=10^{-3}$, adaptive perturbations with $\mu=1$ and  will suppose presence of all types of regularization introduced in section \ref{s2_5}. The Fig. \ref{fig_10}, (a) represents performance curves for the system operating with: no compensation, compensation by SPGD controller with $2\times10^4$ iteration per second and compensation by AI controller having $5\times10^3$ weights updates per second. During training period of 1200 ms the SPGD and AI controllers provide approximately the same compensation quality and corresponding metric curves partially repeat each other except several intervals where AI controller operates better. After turning the training and SPGD iterations off the compensation behavior is dramatically changed -- the SPGD curve partially begins to repeat ``no compensation'' curve whereas AI controller continues to provide the compensation but not so effective as during training phase. 

In the second experiment we will focus on compensation performance of AI controller instead of its learning capabilities  and use more ``aggressive'' training strategy with enhanced learning rate $\gamma_t=10^{-2}$, $\mu=0.5$ and regularization applied to DNN outputs only (i.e. using functional $S_t$ only, see section \ref{s2_5}). As before, the control is synthesized directly using piston/tip-tilt variables. The Fig. \ref{fig_10}, (b) represents numerical results for exactly the same realizations of infinite phase screens as in Fig. \ref{fig_10}, (a). It is easy to see that during training the AI control curve is now lying a little bit higher than SPGD curve and, in general, AI control has less falls in performance in comparison with SPGD control. However, turning the training off leads AI controller to significantly worse compensation in comparison with compensation in Fig. \ref{fig_10}, (a) that can be definitely interpreted as the worse training in this mode.

\begin{figure}[hbtp]
\centering
\includegraphics[width=0.8\textwidth]{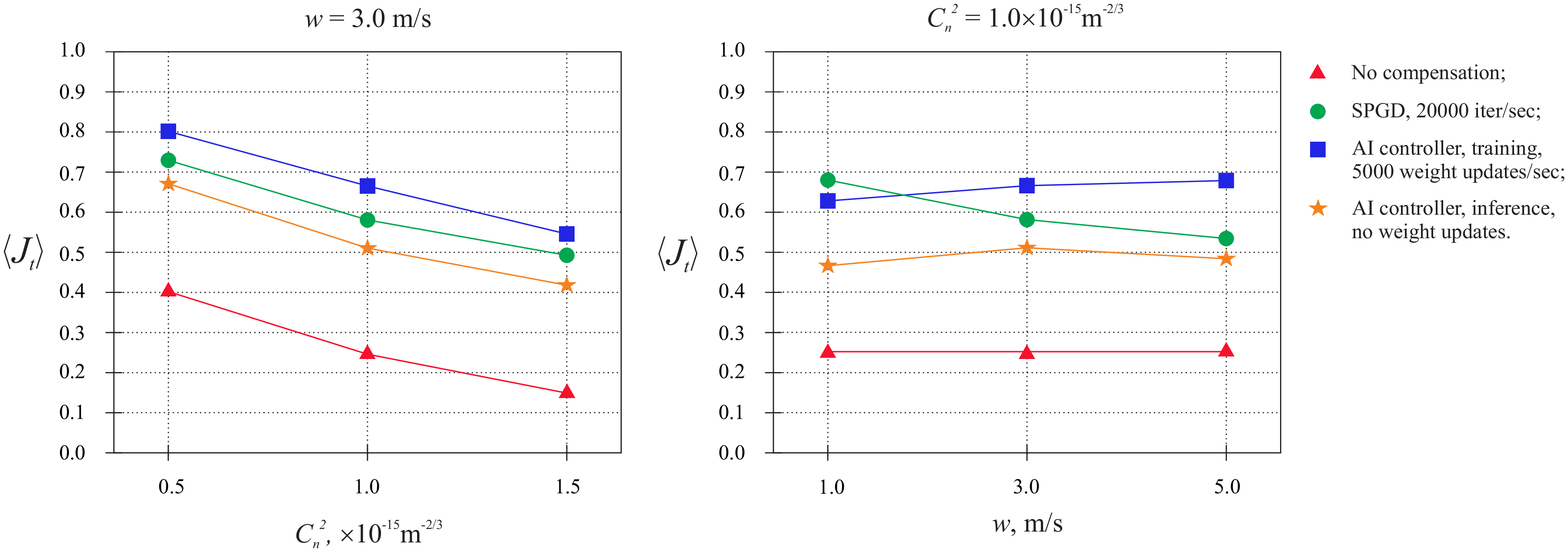}
\caption{Comparison results of SPGD- and AI-based control performance for different atmospheric conditions. Both SPGD and AI controllers operate with piston/tip-tilt control variables, overall simulation time for averaging the results is 5 sec.} 
\label{fig_11}
\end{figure}

The Fig. \ref{fig_11} represents graphical summary of the results obtained via averaging over 5 seconds of corresponding metric values for four compensation strategies mentioned before. The results show that AI controller of proposed topology in training mode exceeds conventional SPGD algorithm up to approximately 5-7\% and in inference mode lose SPGD up to 10\% except slow 1 m/s wind velocity where potentially not a complete training was performed due to slow screens motion.

\section{Conclusions.}
\label{s4}

In this study we introduced self-learning AI controller and applied it for the power beaming problem. The major goal of this research was to verify that controller's DNN can be effectively trained in real time operational regime using solely SPDG-type gradient estimation so that after turning perturbations off the controller keeps a capability to synthesize appropriate control. The numerical results show that this goal can be achieved if during training we use ``soft'' optimization strategy and do not strive to come to maximal metric performance in a minimal time. In opposite, more ``aggressive'' optimization leads to low or even absolutely absence of DNN teaching. The trade off between these two optimization types is empirical and mostly depends on choice of SPGD perturbation statistics, DNN learning rate and its regularization parameters.  

Another important factor that can be emphasized in connection with AI controller learning is a specific organization of training data stream. In the current implementation training process is performed using data portions (of the size $N_{\rm{ws}}$) consequently derived from simulation engine so that DNN weights are updated on each data portion and after that this data is removed. Here we face several problems: (1) the data represented in nearby portions are strongly correlated that has negative effect on training speed, (2) complete removing of used data provides adaptivity of the controller to environment changes but produces instability in training due to factually absence of training dataset, (3) using just one training thread (our batch size equals to 1) leads to essential extension of training process as well as more poor training quality in comparison with multi-thread training. Technically, there is not any obstacle to resolve all these issues however it is a point for a future research. 

In addition to different features of training process it is necessary to note the regularization role in proper controller's functioning. As was mentioned before, applying regularization to DNN outputs as well as smoothing of SPGD gradient allow to prevent such undesirable factor as forming of positive feedback which is typically expressed as generation of shock-type control. On the other hand, excessive regularization leads to non-plasticity in control as well as worsening of DNN training so that here we also have to keep a reasonable trade off. Decoupling of control channels can be also noted as important factor for the controller training and regularization but it is not a surprise due to validity of this observation for a wide class of control systems.

Returning back to power beaming with AI control it is necessary to highlight that DNN's incoming information in the form of PVA sensor data in principle does not contain all required information for synthesis optimal and unique phase pistons and tip-tilts. The lack of this information can be filled or passing additional sensor data to DNN input or including constantly working optimization mechanism like SPGD or DNN weights updating. In the last case the question how long DNN control will be stable during such a continuous training is still open.

In conclusion note that the proposed concept of active AI control potentially can be extended to a wide class of physical systems where SPGD provides appropriate estimation of metric gradient. Presence of metric singular points over control variables or in systems with strong inertia over control this approach should be corrected in the part of proper metric gradient estimation. However, in presence of boundary conditions and other strict limitations in control space and in systems where optimization process requires complicated predictive policy this approach, apparently, is not so effective and $Q$-function approach is preferable.    

\section{Acknowledgment.}
\label{s5}
The authors are sincerely grateful to M.A. Vorontsov for useful discussions of fiber-array power beaming problem and A.B. Lavrentyev for the interest in AI-based control systems.


\end{document}